\documentclass[acmtog, nonacm]{acmart}
\acmSubmissionID{683}
\usepackage{booktabs} % For formal tables

\setcitestyle{nosort,square} % nosort to allow for manual chronological ordering

\acmJournal{TOG}
\usepackage[separate-uncertainty=true, per-mode=fraction, alsoload=synchem]{siunitx}[=v2]
\usepackage[version=4]{mhchem}
\usepackage{algorithm}
\usepackage{algpseudocode}
\usepackage{tabularx}

\usepackage{xargs}
\usepackage[normalem]{ulem}
\newcommandx{\added}[1]{#1}
\newcommandx{\removed}[1]{}
\newcommandx{\replaced}[2]{\removed{#1}\added{#2}}

\begin{document}

\title[Single-View Holographic Volumetric 3D Printing (SHVAM)]{Single-View Holographic Volumetric 3D Printing with Coupled Differentiable Wave-Optical and Photochemical Optimization}

\author{Felix Wechsler}
\authornotemark[1]
\affiliation{%
 \institution{\'Ecole Polytechnique Fédérale de Lausanne (EPFL)}
\country{Switzerland}}
\email{shvam@felixwechsler.science}

\author{Riccardo Rizzo}
\affiliation{%
\institution{\'Ecole Polytechnique Fédérale de Lausanne (EPFL)}
\country{Switzerland}}
\email{riccardo.rizzo@epfl.ch}

\author{Christophe Moser}
\affiliation{%
\institution{\'Ecole Polytechnique Fédérale de Lausanne (EPFL)}
\country{Switzerland}}
\email{christophe.moser@epfl.ch}

\renewcommand\shortauthors{Wechsler et al.}


\begin{abstract}
Volumetric additive manufacturing promises near-instantaneous fabrication of 3D objects, yet achieving high fidelity at the micro-scale remains challenging due to the complex interplay between optical diffraction and chemical effects. We present \emph{Single-View Holographic Volumetric Additive Manufacturing} (SHVAM), a mechanically static system that shapes volumetric dose distributions using time-multiplexed, phase-only holograms projected from a single optical axis. To achieve high resolution with SHVAM, we formulate hologram synthesis as a coupled inverse problem, integrating a differentiable wave-optical forward model with a simplified photochemical model that explicitly captures inhibitor diffusion and non-linear dose response. Optimizing hologram sequences under these coupled constraints allows us to pre-compensate for chemical blur, yielding higher print fidelity than optical-only optimization. We demonstrate the efficacy of SHVAM by fabricating simple 2D and 3D structures with lateral feature sizes of approximately \SI{10}{\micro\meter} within a $\SI{0.8}{\milli\meter} \times \SI{0.8}{\milli\meter} \times \SI{3}{\milli\meter}$ volume in seconds.
\end{abstract}



%\keywords{Holography, 3D printing, volumetric additive manufacturing, wave optics, differentiable optics}
\begin{teaserfigure}
  \includegraphics[width=\textwidth]{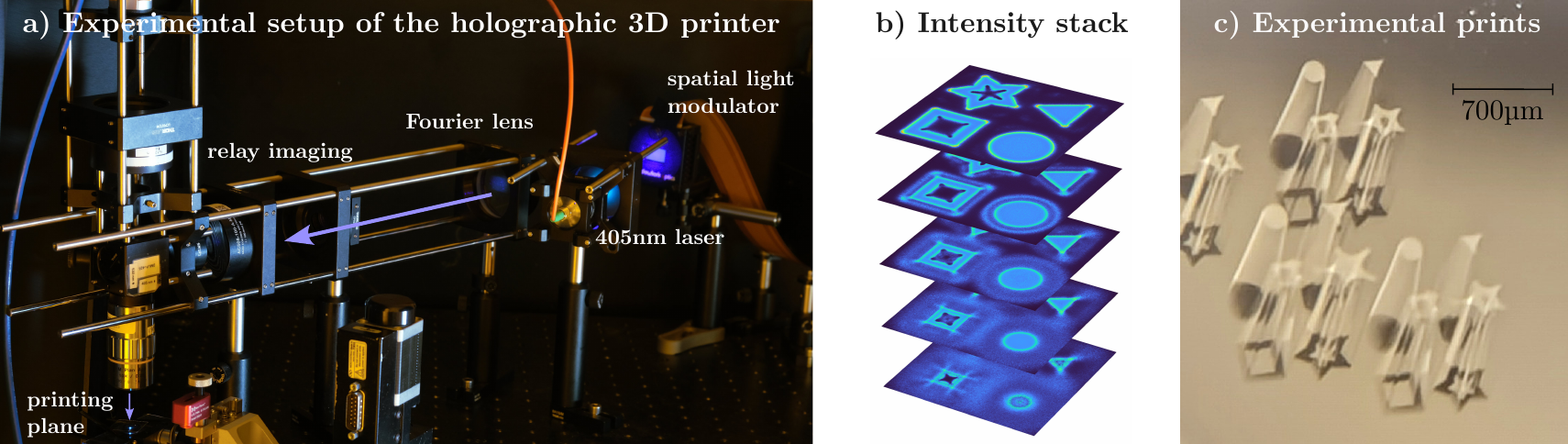}
  \caption{\textbf{a)} SHVAM overview: a \SI{405}{\nano\meter} laser is phase-modulated by a spatial light modulator (SLM) and relayed through a Fourier holography optical system into a photosensitive resin, enabling mechanically static, single-axis volumetric exposure. 
  \textbf{b)} Volumetric printing is posed as a coupled inverse problem: we optimize a time-multiplexed sequence of phase-only holograms so that the \emph{accumulated} 3D dose matches a target, while explicitly accounting for inhibition and diffusion in a differentiable photochemical model to pre-compensate chemical blur. 
  \textbf{c)} Representative prints demonstrating rapid fabrication and improved fidelity from chemistry-aware optimization; each group of elements is fabricated in \SI{8}{\second}.}
\end{teaserfigure}


\maketitle

    \section{Introduction}

    To overcome fundamental speed limitations of layer-by-layer printing while retaining geometric freedom, Volumetric Additive Manufacturing (VAM) was introduced to fabricate entire objects in a single exposure process \cite{Shusteff_Browar_Kelly_Henriksson_Weisgraber_Panas_Fang_Spadaccini_2017}. In Tomographic Volumetric Additive Manufacturing (TVAM), a 3D volume is formed through tomographic back-projection of 2D patterns while a resin vial rotates \cite{kelly2019volumetric, bernal2019volumetric, Loterie_Delrot_Moser_2020}. 

    A key challenge across these lithographic techniques is that print outcomes depend not only on optics but also on resin chemistry. Accurate chemistry modeling is essential to predict reaction kinetics and diffusion processes that govern polymerization and final material properties \cite{weisgraber2023virtual}. In particular, diffusion of chemical inhibitors such as oxygen or TEMPO (2,2,6,6-tetramethylpiperidine 1-oxyl) can produce visible defects if not accounted for \cite{Orth_Webber_Zhang_Sampson_de,Zhang_DeHaan_Houlahan_Sampson_Webber_Orth_Lacelle_Gaburici_Lam_Deore_etal._2025}. Data-driven approaches can learn systematic correction of prints, as demonstrated for Two-Photon polymerization \cite{neurallithography}.

    In this work, we focus on volumetric printing driven by computer-generated holography (CGH), which enables shaping optical fields by phase modulation. A common implementation uses a Spatial Light Modulator (SLM) to modulate a coherent beam in Fourier space, with a lens mapping phase modulation to intensity in the target plane; off-axis carrier ramps are often used to separate the reconstruction from the zero-order spot \cite{slmbook, goodman2017introduction}. Recent advances in CGH increasingly rely on differentiable wave propagation models to improve fidelity and compensate for system aberrations \cite{ 10.1145/3414685.3417802, 10.1145/3478513.3480542}.

    \begin{figure}
        \centering
        \includegraphics[width=.475\textwidth]{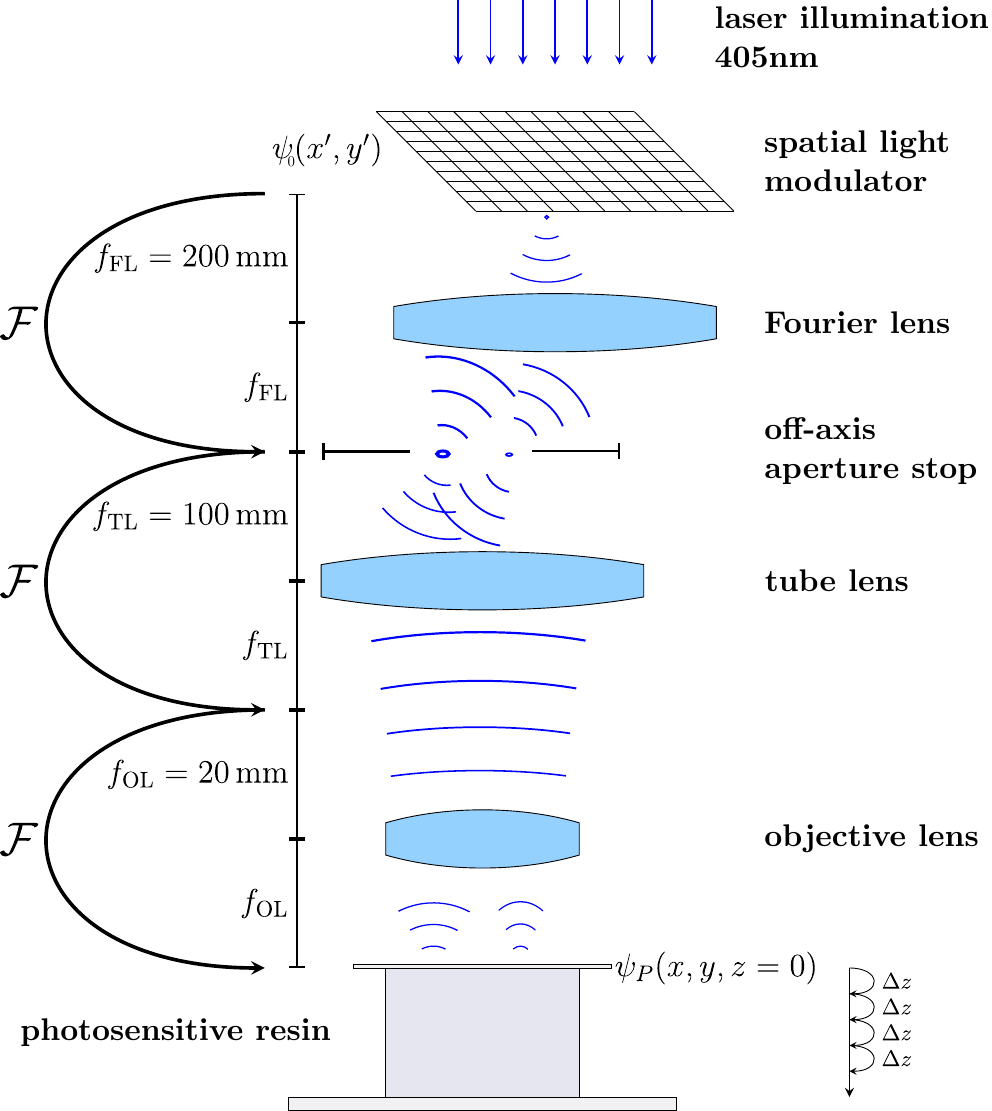}
        \caption{Schematic of the SHVAM printer. A coherent beam is phase-modulated by a spatial light modulator (SLM) and Fourier-transformed by a lens. The strong zero-order (unmodulated) component arising from the pixelated SLM is suppressed using an off-axis spatial filter. A relay imaging system (tube lens and microscope objective) then re-images the filtered field into the resin at the printing plane ($z=0$), from which the 3D field distribution is computed via angular-spectrum wave propagation.}
        \label{fig:setup}
    \end{figure}

    We introduce a mechanically static, single-view volumetric printing setup that uses Fourier-domain holographic projections to fabricate 3D objects (see \autoref{fig:setup}). We call our approach \emph{Single-View Holographic Volumetric Additive Manufacturing} (SHVAM). SHVAM optimizes a time-multiplexed sequence of phase-only holograms such that the accumulated 3D energy deposition in a photosensitive resin matches a target dose distribution. In our prototype, we project holograms using a coherent \SI{405}{\nano\meter} laser and a phase-only SLM imaged through a microscope objective.
    This enables faster VAM printing, as it eliminates mechanical rotation and relies solely on the laser power and projection refresh rate.
    
    Producing complex volumetric dose distributions from a single optical axis requires solving a constrained inverse problem. We therefore couple phase optimization with a wave-optical forward model based on angular-spectrum propagation. To improve fidelity for fine features, we further incorporate a photochemical model directly into the objective \added{loss function}, accounting for radical generation, inhibitor concentration, and inhibitor diffusion during exposure. We validate the proposed model and optimization by fabricating a range of 2D and 3D structures in seconds and by comparing predictions against experimental outcomes.

    Our contributions are:
    \begin{itemize}
        \item A single-view, phase-only hologram synthesis method for volumetric 3D printing based on a wave-optical forward model.
        \item A differentiable photochemical model for volumetric 3D printing that explicitly captures inhibitor diffusion and non-linear dose response. While demonstrated here for holographic printing, this formulation is applicable to many light-based volumetric printers (e.g., TVAM).
        \item Experimental validation on 2D and 3D prints, demonstrating the impact of the photochemical model on print fidelity.
    \end{itemize}

   We discuss current limitations in printable volume and fidelity, and outline directions to scale SHVAM through improved calibration, optics, and resin characterization. By adopting a static single-view architecture instead of mechanical rotation (as in TVAM), we explicitly trade axial resolution for mechanical simplicity and fabrication speed. While the axial resolution remains limited by the low numerical aperture (NA) required to maintain a wide field of view, this configuration enables the rapid generation of micro-structures without moving parts. Our reference implementation can be found at: \url{https://github.com/EPFL-LAPD/SHVAM}

\section{Background and related work}
\paragraph{Additive Manufacturing}
   Additive manufacturing has evolved from rapid prototyping into a family of processes capable of in-dustrial-scale and micro-scale fabrication. Classic layer-by-layer approaches include extrusion-based methods such as Fused Deposition Modeling \cite{crump1992apparatus} and vat photopolymerization techniques such as Stereolithography (SLA) \cite{hull1986apparatus, Kodama_1981} and Digital Light Processing \cite{hornbeck1997digital}. A major step toward faster photopolymer printing was Continuous Liquid Interface Production, which uses an oxygen-permeable window to form a \textit{dead zone} that enables continuous fabrication \cite{tumbleston2015continuous}. In parallel, the demand for high-resolution microfabrication led to micro-SLA \cite{Ikuta_Hirowatari_1993} and, for sub-diffraction precision, Two-Photon Polymerization (2PP) \cite{maruo1997three}. While 2PP offers exceptional resolution, volumetric throughput is typically limited by serial point scanning; holographic multiplexing can increase throughput by generating multiple foci simultaneously \cite{ganni2024holographic}, and more complex holographic 2PP has been theoretically considered for high-power single laser pulses \cite{Somers:24}.

\paragraph{Volumetric Additive Manufacturing (VAM)}
    SHVAM is most closely connected to prior work on volumetric additive manufacturing. Early volumetric fabrication demonstrations (not yet referred to as VAM) used holographic interference of coherent beams to create periodic structures \cite{Campbell_Sharp_Harrison_Denning_Turberfield_2000}. Later, \citet{Shusteff_Browar_Kelly_Henriksson_Weisgraber_Panas_Fang_Spadaccini_2017} introduced a VAM system using three intersecting (but not coherently interfering) beams. This formulation admits efficient fabrication but restricts printable geometries because the light transport is modeled as a small set of collimated rays, limiting achievable shapes to combinations compatible with the available beam directions.

\paragraph{Tomographic Volumetric Additive Manufacturing (TVAM)}
    Tomographic Volumetric Additive Manufacturing (TVAM) generalizes VAM by projecting a sequence of tomographic light patterns into a rotating vial of photosensitive resin \cite{kelly2019volumetric, Loterie_Delrot_Moser_2020}. In these systems, projected energy is attenuated according to Beer--Lambert absorption \cite{kelly2019volumetric} and initiates free-radical polymerization. TVAM relies on a nonlinear threshold response: projected patterns are optimized so that target regions exceed the polymerization threshold while void regions remain below it \cite{kelly2019volumetric,Rackson_Champley_Toombs_Fong_Bansal_Taylor_Shusteff_McLeod_2021,BHATTACHARYA2021102299}. Inverse rendering techniques that explicitly model light transport enable improved fidelity across optical scenarios \cite{Nicolet_Wechsler_Madrid-Wolff_Moser_Jakob_2024}. TVAM is also particularly attractive for bioprinting \cite{bernal2019volumetric} and overprinting onto existing absorbing, scattering, refracting or reflective structures \cite{wechsler2025overprintingtomographicvolumetricadditive}.

\paragraph{Chemical effects in VAM}
    A practical complication is inhibition. Once radicals are generated, they can be quenched by inhibitors such as dissolved oxygen; polymerization initiation and propagation become effective only after local inhibitor depletion \cite{Zhang_DeHaan_Houlahan_Sampson_Webber_Orth_Lacelle_Gaburici_Lam_Deore_etal._2025}. This depletion can introduce spatiotemporal dynamics, including diffusion-driven blur that affects print fidelity. 
    \added{Beyond oxygen, TEMPO has been explored as an additional inhibitor \cite{Toombs_Luitz_Cook_Jenne_Li_Rapp, THIJSSEN2024106096};} 
    \citet{Orth_Webber_Zhang_Sampson_de} modeled \replaced{inhibitor}{oxygen}-related artifacts by modifying the distributed dose with a convolution-like operation, which captures some qualitative \added{blurring} effects but does not explicitly represent chemical diffusion in the resin. 
    \added{Their approach does not account for the finite oxygen concentration or its local depletion to zero within polymerized regions. Consequently, it describes the evolving oxygen concentration field inaccurately and effectively applies diffusion-induced blurring of the light dose throughout the entire printing process. Moreover, it does not readily accommodate multiple inhibitor species (such as oxygen and TEMPO) with specific diffusion coefficients and initial concentrations.}
    \removed{Beyond oxygen, TEMPO has been explored as an additional inhibitor [Toombs et al. 2022; Thijssen et al. 2024];} 
    \added{TEMPO related} improved fidelity is often attributed to reduced diffusion compared to oxygen, though quantitative integration of TEMPO dynamics into pattern generation remains limited.

\paragraph{Wave-optical TVAM projections}
    On the optical side, TVAM patterns are most commonly projected via amplitude modulation, but phase modulation has also been demonstrated \cite{holotvam}. That work, however, uses a ray-optical description and therefore does not constitute a fully wave-optical holographic model of light transport. \citet{Wechsler_Gigli_Madrid-Wolff_Moser_2024} introduced a theoretical wave-optical light transport formulation for TVAM, which is, in principle, compatible with phase-modulated systems.
    
\paragraph{Holographic VAM}
    Related wave-optical optimization schemes for holographic printers have been presented by \citet{Li:24} and \citet{MoserWechslerAlvarezCastano2025WO2025223658A1}; \citet{Li:24} additionally modeled an inhibition beam, and later provided simple experimental demonstrations with two orthogonal projections \cite{li2024developments}. The use of multiple views (compared to one view in this work) was motivated by the limited spatial-frequency coverage of a single projection objective (see \autoref{sec:missingcone}). Very recently, a static single-view phase-mask-based holographic printer was proposed \cite{lin2026singleexposureholographic3dprinting}. While closely related in basic principles to this work, it relies on a single, non-reconfigurable phase element, producing one coherent field realization in the volume and thus remaining constrained by monochromatic wave propagation (the Helmholtz equation). In contrast, SHVAM uses time-multiplexed phase patterns whose dose accumulation is effectively incoherent (when a large amount of patterns is used), enabling volumetric dose distributions that are not attainable from any single coherent propagation; see \autoref{sec:multiple_patterns} for discussion. Further, their work does not model any chemical effects quantitatively and will result in limited resolutions as we demonstrate in this work.

\section{Methodology}
    
    The core of SHVAM's modeling is a wave-optical simulation that describes how a phase-modulated coherent pattern produces a 3D intensity distribution. The accumulated polymerization dose in the photosensitive resin results from time-integrating the intensities from a set of such patterns.
    However, as shown in this work, additionally modeling the photochemical kinetics of the resin is key to achieving high-resolution prints. We therefore introduce explicit (albeit simplified) modeling of inhibitors and their chemically induced diffusion.

\subsection{Optical model}
    The optical setup is shown in \autoref{fig:setup}. We use a phase-modulating spatial light modulator (SLM) to imprint a phase pattern $\varphi_j$ onto an incident coherent field $\psi_0(x',y')$. The modulated field is relayed through a standard Fourier holography configuration: a Fourier lens of focal length $f_\text{FL}$ performs an optical Fourier transform, and an off-axis aperture selects a single diffraction order to suppress the strong on-axis (zero-order) component caused by unmodulated light from the SLM \cite{slmbook}. The selected field is then reimaged into the resin volume using a tube lens and a microscope objective. The overall demagnification from the aperture plane to the printing plane is
\begin{equation}
    M = \frac{f_\text{OL}}{f_\text{TL}} \, ,
\end{equation}
where $f_\text{OL}$ and $f_\text{TL}$ denote the focal lengths of the objective and tube lens, respectively.

Let $\Delta x$ be the SLM pixel pitch and $N$ the number of SLM pixels along one dimension. Under the usual Fourier scaling for a lens-based Fourier transform \cite{goodman2017introduction}, the effective sampling in the aperture (Fourier) plane is
\begin{equation}
    \Delta x_{A} = \frac{\lambda f_\text{FL}}{N \Delta x} \, .
\end{equation}
After demagnification into the printing plane, the effective sampling becomes
\begin{equation}
    \Delta x_{P} = \frac{\lambda f_\text{FL} f_\text{OL}}{N \Delta x\, f_\text{TL}} \, ,
\end{equation}
where $\lambda=\SI{405}{\nano\meter}$ is the vacuum wavelength.

Given a displayed phase pattern $\varphi_j(x',y')$ on the SLM, the complex field at the printing plane ($z=0$) can be written (up to a constant scaling factor) as
\begin{equation}
    \psi_{P, j}(x,y,0) \sim
    \mathcal{F}\!\left[
        \psi_0(x',y') \,\exp\!\left(i\,\varphi_j(x',y')\right)
    \right](x,y),
\end{equation}
where $(x',y')$ are continuous coordinates on the SLM and $(x,y)$ denote continuous coordinates in the printing plane.

To model propagation from the printing plane into the resin volume, we use band-limited angular spectrum propagation \cite{Matsushima_Shimobaba_2009, Wechsler_Gigli_Madrid-Wolff_Moser_2024}:
\begin{equation}
    \psi_{P, j}(x,y,z) =
    \mathcal{F}^{-1}\!\Big[
        H_\text{AS}(z)\,\mathcal{F}\!\left[\psi_{P, j}(x,y,0)\right]
    \Big]
    \;=\;
    \mathcal{A}_z[\psi_{P,j}(x,y,0)] \, ,
\end{equation}
with transfer function
\begin{equation}
    H_\text{AS}(z) =
    \exp\!\left(
        i\,2\pi n z
        \sqrt{\frac{1}{\lambda^2} - f_x^2 - f_y^2}
    \right),
\end{equation}
where $n$ is the refractive index of the resin and $(f_x,f_y)$ are spatial frequencies.

In the context of printing, the resin integrates optical intensity over time, and light is attenuated along depth according to Beer--Lambert absorption. We therefore model the accumulated absorbed dose $D(x,y,z)$ as
\begin{equation}
\begin{split}
    D(x,y,z)
    \sim
    T_\text{exp}\,
    \mu\,
    \exp(-\mu z)\,
    \sum_{j=1}^{K}
    \left|
        \mathcal{A}_z\!\left[
            \psi_{P,j}(x,y,0)
        \right]
    \right|^2,
\end{split}
\label{eq:wavefull}
\end{equation}
where $\mu$ is the absorption coefficient of the resin (primarily determined by photoinitiator concentration), $T_\text{exp}$ is the total exposure time, $K$ is the number of time-multiplexed phase patterns, and $\psi_{P,j}$ denotes the printing-plane field produced by phase pattern $\varphi_j$.

We sum over patterns because each hologram is displayed for a short interval $\Delta t \ll T_\text{exp}$ and the resin effectively accumulates the \emph{incoherent} sum of intensities. Time multiplexing provides two practical benefits: (i) it reduces coherent speckle artifacts through temporal averaging, and (ii) it increases the expressiveness of attainable 3D dose distributions compared to a single coherent field realization (see \autoref{sec:multiple_patterns}). Equation~\ref{eq:wavefull} is fully differentiable and can be evaluated with FFT-based propagation at a computational cost of approximately $\mathcal{O}(K\cdot N_z \cdot N^2\log N)$ for a cubic $N\times N\times N_z$ voxel volume.

    \begin{figure}
        \centering
        \includegraphics[width=1\linewidth]{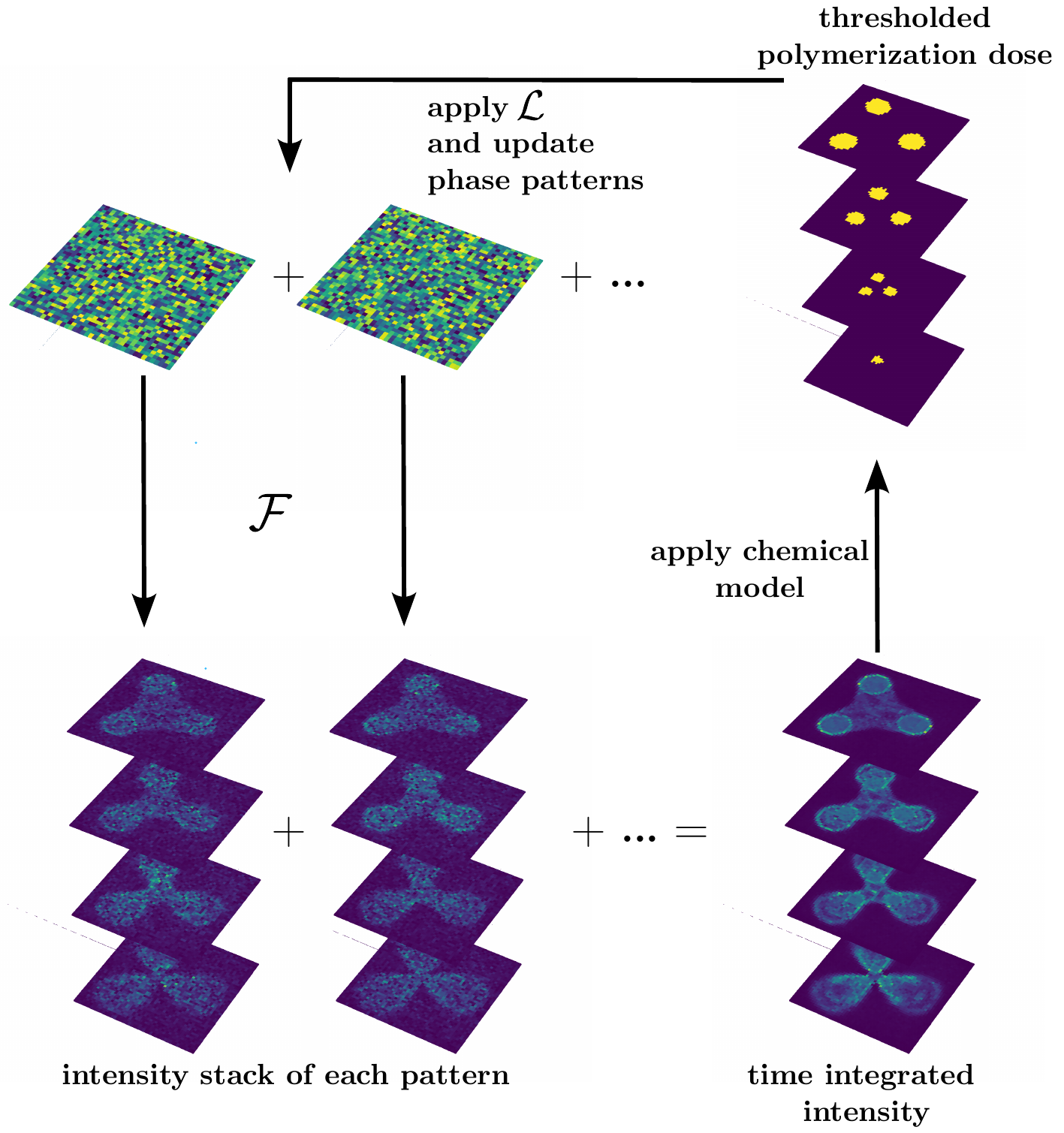}
        \caption{Visual description of the optimization process. A fixed number of phase patterns is projected into the resin which results in the intensity summation of those patterns. Their sum results in a time integrated intensity. After applying the chemical model we obtain a binary solidified print. Evaluating this print and the chemical concentrations with our loss function allows to update the phase patterns via gradient-descent based optimization.}
        \label{fig:schematics}
    \end{figure}

\subsection{Chemical model}
    
    The second component of our forward model is a simplified description of the underlying photochemistry. 
    Given the wave-optical intensity field in the resin volume, the chemical model predicts a polymerization field and thus a final (thresholded) print outcome, as illustrated in \autoref{fig:schematics}.

    Although we apply this model to a single-view holographic setup, the underlying reaction-diffusion kinetics depend solely on the accumulated 3D light intensity distribution and material parameters. Consequently, this differentiable chemical module is agnostic to the optical delivery method and can be readily adapted to improve fidelity in other volumetric approaches.
    
    In principle, the full reaction network governing volumetric additive manufacturing (VAM) can be modeled in detail \cite{weisgraber2023virtual}. However, comprehensive kinetic models require characterization of a large set of rate constants and material parameters and are typically used as forward simulators rather than as differentiable components inside an inverse-design loop. To avoid an extensive parameter-identification procedure while still capturing the dominant resolution-limiting mechanisms, we adopt a simplified model that retains the key effects of inhibition and inhibitor diffusion.
    
    In TVAM and related approaches, polymerization is approximated by a threshold model: a light pattern is projected for a prescribed duration, and voxels are assumed to solidify once the absorbed dose exceeds a fixed threshold. While convenient, this description is not sufficiently accurate for predicting (and compensating) inhibition-driven effects.
    
    At the onset of exposure, incident light is absorbed by the photoinitiator (PI), generating reactive radicals that can initiate polymerization. In practice, inhibition reactions due to dissolved oxygen and/or added inhibitors such as TEMPO are orders of magnitude faster than propagation reactions \cite{weisgraber2023virtual, Zhang_DeHaan_Houlahan_Sampson_Webber_Orth_Lacelle_Gaburici_Lam_Deore_etal._2025}. Consequently, as long as the local inhibitor concentration is non-zero, radical quenching dominates and polymerization is strongly suppressed. Once inhibitors are locally depleted, concentration gradients form and induce diffusive transport, which couples the chemistry across space.
    
    Our resin contains two inhibitors: oxygen and TEMPO. The latter is comparatively large and diffuses slowly, whereas oxygen diffuses significantly faster. Starting from the full model in \cite{weisgraber2023virtual}, we apply the following assumptions to obtain a computationally tractable reduced model:
    
    \begin{itemize}
        \item The monomer concentration $c[\ce{M}]$ is assumed constant over the exposure time.
        \item Radicals that are not inhibited contribute to an accumulated polymerization field $c[\ce{P}^*]$.
        \item The temperature is constant.
        \item Polymerization is local (i.e., we neglect spatial growth and transport of polymer chains).
        \item Diffusion of inhibitors (\ce{O2}, TEMPO) is modeled explicitly; diffusion of PI and radicals is neglected due to their comparatively low mobility (consistent with the size dependence of diffusion described by the Stokes--Einstein relation).
    \end{itemize}
    
    Note that PI depletion in principle occurs during illumination; including it would require dynamically updating local absorption and increases computational cost. In our reduced model, we therefore treat PI as effectively constant and absorb its contribution into an effective radical-generation term. Likewise, we do not explicitly model chain growth and termination kinetics; instead, their net effect is captured by an effective propagation term contributing to $c[\ce{P}^*]$.
    
    Under these assumptions we can analyze the kinetic rate equations:
    \begin{align*}
    \frac{\partial c[\ce{R}]}{\partial t} &=
    2\,k_I\,I\,c[\ce{PI}]
    - k_p\,c[\ce{M}]\,c[\ce{R}]
    - k_{\ce{O2}}\,c[\ce{O2}]\,c[\ce{R}]\\
    &-k_{\ce{TEMPO}}\,c[\ce{TEMPO}]\,c[\ce{R}] \\
    \frac{\partial c[\ce{O2}]}{\partial t} &=
    \nabla^2 \cdot \left(D_{\ce{O2}}\, c[\ce{O2}]\right)
    - k_{\ce{O2}}\,c[\ce{O2}]\,c[\ce{R}] \\
    \frac{\partial c[\ce{TEMPO}]}{\partial t} &=
    \nabla^2 \cdot \left(D_{\ce{TEMPO}}\, c[\ce{TEMPO}]\right)
    - k_{\ce{TEMPO}}\,c[\ce{TEMPO}]\,c[\ce{R}] \\
    \frac{\partial c[\ce{P}^*]}{\partial t} &=
    k_p\,c[\ce{M}]\,c[\ce{R}] \, .
    \end{align*}
    Here, $I$ denotes the local light intensity, $k_I$ is the rate constant for light-triggered radical generation, $k_p$ is an effective propagation rate constant, $k_{\ce{O2}}$ and $k_{\ce{TEMPO}}$ govern inhibition, and $D_{\ce{O2}}, D_{\ce{TEMPO}}$ are diffusion coefficients.
    
    For numerical efficiency in inverse design, we exploit the separation of timescales $k_{\ce{O2}} > k_{\ce{TEMPO}} \gg k_p$, which implies that newly generated radicals are preferentially quenched by oxygen, then by TEMPO, and only the remaining radicals contribute substantially to polymerization. 
    We therefore do not explicitly integrate the reaction--diffusion system since it would require precise knowledge of the absolute values of photochemical constants. Instead, we use an operator-splitting surrogate: at each time step we (i) generate radicals proportional to the local intensity, (ii) quench radicals by available inhibitors (oxygen, then TEMPO), (iii) accumulate polymerization from the remaining radicals, and (iv) diffuse inhibitor concentrations.
    
    Finally, the inhibitor diffusion step can be implemented efficiently using \removed{the} Green's function of the diffusion equation. For constant diffusion coefficient $D$, a diffusion step of duration $\Delta T$ corresponds to convolution with a Gaussian kernel,
    \begin{equation}
    K^{\Delta T}(x,y,z) = \frac{1}{(4\pi D \Delta T)^{3/2}} \exp\!\left(-\frac{x^2+y^2+z^2}{4D \Delta T}\right),
    \end{equation}
    which we evaluate numerically via FFT-based convolutions \cite{Orth_Webber_Zhang_Sampson_de}.
    \added{Green's function to model diffusion is only valid for free space. Our printing region can be considered as small compared to the overall resin region}.
    We do not apply padding (usually applied to avoid circular convolution artifacts) as our kernel size is small and we have sufficient padding around the target \added{such that we can ignore wrapping artifacts}.
    Overall, the reduced chemical modeling used inside our inverse-design loop is summarized in Algorithm \ref{alg:chem}.
    To connect the time-discretized chemistry surrogate to the experiment, we interpret one chemical timestep $\Delta T$ as one \emph{pattern cycle}: the duration required to display the full set of $K$ optimized holograms once on the SLM (each for $\Delta t$, with $\Delta T = K\,\Delta t$). During a print, this cycle is repeated $N_{\text{rep}}$ times, so that the total exposure time is $T_\text{exp} = N_{\text{rep}}\,\Delta T$ (in our experiments, $N_{\text{rep}}\approx 15$--$20$). Within each cycle, we model radical generation using the intensity produced by the $K$ holograms (Eq.~\ref{eq:wavefull}), while inhibitor diffusion and depletion are updated once per cycle via the chemistry loop.\\
    Despite the strong simplifications of this approach, we show in \autoref{sec:results} that incorporating inhibitor diffusion inside the inverse-design loop yields consistent practical improvements in print fidelity.
    
    \begin{algorithm}
    \caption{Reduced photochemistry surrogate with inhibition and diffusion (used inside inverse design)}
    \label{alg:chem}
    \begin{algorithmic}[1]
    \State \textbf{Input:} target design, time step $\Delta T$, total exposure time $T_\text{exp}$, diffusion kernels $K^{\Delta T}_{\ce{O2}}, K^{\Delta T}_{\ce{TEMPO}}$, scaling parameter $\alpha$
    \State \textbf{Output:} Phase patterns $\{\varphi_j\}_{j=1}^K$
    \State $N_T \gets  T_\text{exp} / \Delta T$
    \Statex
    
    \For{optimization iteration $m = 1,2,\dots,M$}
        \State $c[\ce{P}^*] \gets 0$
        \State $c[\ce{O2}] \gets c[\ce{O2}]_0$
        \State $c[\ce{TEMPO}] \gets c[\ce{TEMPO}]_0$
        \State Compute dose field $D(\mathbf r)$ from $\{\varphi_j\}$ (Eq.~\ref{eq:wavefull})
        \Statex
        \For{$n = 1,2,\dots,N_T$}
 
            \State \textbf{Radical generation:} $c[\ce{R}] \gets \alpha \, D / N_T$
            \State \textbf{Quenching oxygen:} 
            \State $q_{\ce{O2}} \gets \min\!\left(c[\ce{R}],\, c[\ce{O2}]\right)$
            \State $c[\ce{R}] \gets c[\ce{R}] - q_{\ce{O2}}$
            \State $c[\ce{O2}] \gets c[\ce{O2}] - q_{\ce{O2}}$
            \State \textbf{Quenching TEMPO:}
            \State $q_{\ce{TEMPO}} \gets \min\!\left(c[\ce{R}],\, c[\ce{TEMPO}]\right)$
            \State $c[\ce{R}] \gets c[\ce{R}] - q_{\ce{TEMPO}}$
            \State $c[\ce{TEMPO}] \gets c[\ce{TEMPO}] - q_{\ce{TEMPO}}$
            \State \textbf{Polymerization:} $c[\ce{P}^*] \gets c[\ce{P}^*] + c[\ce{R}]$
            \State{\# $\ast$ \textit{indicates convolution}}
            \State \textbf{Diffusion:}
            $c[\ce{O2}] \gets c[\ce{O2}] \ast K^{\Delta T}_{\ce{O2}}$
            \State \hspace{1.45cm}
            $c[\ce{TEMPO}] \gets c[\ce{TEMPO}] \ast K^{\Delta T}_{\ce{TEMPO}}$
        \EndFor
    
        \State Compute loss $\mathcal{L}\!\left(c[\ce{P}^*], c[\ce{O2}], c[\ce{TEMPO}]\right)$ (Eq.~\ref{eq:loss})
        \State Update $\{\varphi_j\}$ using L-BFGS
    \EndFor
    \end{algorithmic}
    \end{algorithm}

\subsection{Inverse modeling with chemically-informed loss function}

To compute a set of holographic projection patterns, we solve a differentiable inverse problem over the SLM phase patterns $\{\varphi_j\}_{j=1}^K$. We use a gradient-based optimizer (L-BFGS \cite{Liu_Nocedal_1989}) and backpropagate through the coupled forward model consisting of (i) wave-optical propagation and time-multiplexed intensity accumulation (Eq.~\ref{eq:wavefull}) and (ii) the reduced photochemistry surrogate (Algorithm \ref{alg:chem}). As in other VAM approaches, the goal is to trigger sufficient polymerization in object regions while suppressing polymerization in void regions.

Let $\Omega_{\text{obj}}$ denote the set of object voxels and $\Omega_{\text{void}}$ the set of void voxels. The chemical forward model outputs the accumulated polymerization field $c[\ce{P}^*](\mathbf r)$ as well as inhibitor fields $c[\ce{O2}](\mathbf r)$ and $c[\ce{TEMPO}](\mathbf r)$ at the end of exposure. We define a loss that (i) enforces a target polymerization range in object voxels, (ii) penalizes inhibitor remaining inside the object (encouraging depletion where polymerization should occur), (iii) suppresses polymerization in void voxels, and (iv) enforces that void regions retain a minimum amount of inhibitor, which empirically mitigates unintended curing due to inhibition loss.

We use $\mathrm{ReLU}(x)=\max(0,x)$ and define the loss:
\begin{equation}
\begin{split}
\mathcal{L} =
&\;
\sum_{v \in \Omega_{\text{obj}}}
\left|\mathrm{ReLU}\!\left(T_P - c[\ce{P}^*]_v\right)\right|^2
+
\left|\mathrm{ReLU}\!\left(c[\ce{P}^*]_v - T_{\text{OP}}\right)\right|^2
\\
&\; +
\sum_{v \in \Omega_{\text{obj}}}
\left(
\left|c[\ce{O2}]_v\right|^2
+
\left|c[\ce{TEMPO}]_v\right|^2
\right)
\\
&\; +
\sum_{v \in \Omega_{\text{void}}}
\left(c[\ce{P}^*]_v\right)^2
\\
&\; + 
\sum_{v \in \Omega_{\text{void}}}
\left|
\mathrm{ReLU}\!\left(T_I - c[\ce{O2}]_v - c[\ce{TEMPO}]_v\right)
\right|^2 .
\end{split}
\label{eq:loss}
\end{equation}

Here, $T_P$ is the minimum polymerization level required for solidification (we use $T_P=0.1$), $T_{\text{OP}}$ is an over-polymerization threshold used to discourage excessive curing in object regions (we use $T_{\text{OP}}=0.3$), and $T_I$ is a minimum inhibitor level desired in void regions (we use $T_I=0.2$). Relative weights of the terms can be introduced but were not necessary for our experiments.

\begin{figure*}[t]
    \centering
    \includegraphics[width=1\textwidth]{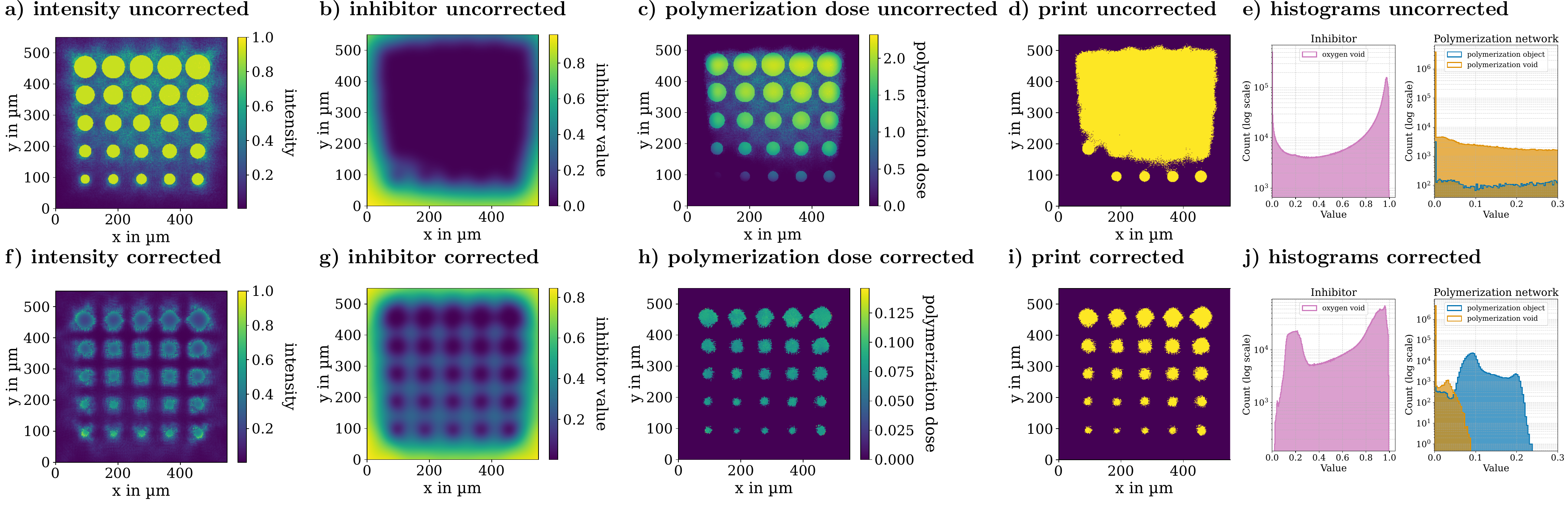}
    \caption{Impact of oxygen diffusion on inverse-designed dose patterns. We compare holograms optimized \emph{without} diffusion (top row) and \emph{with} diffusion-aware optimization (bottom row), using $D_{\ce{O2}}=\SI{200}{\micro\meter\squared\per\second}$. \textbf{a,f)} projected intensity (slice), \textbf{b,g)} oxygen concentration after exposure, \textbf{c,h)} accumulated polymerization field, \textbf{d,i)} thresholded print prediction, and \textbf{e,j)} histograms of oxygen (pink) and polymerization values for object (blue) and void (orange) voxels. Neglecting diffusion depletes oxygen throughout the volume and reduces class separation, leading to unintended curing in void regions; diffusion-aware optimization preserves inhibitor in voids and yields a cleaner thresholded reconstruction.}
    \label{fig:diffusion_impact}
\end{figure*}

In all subsequent experiments, we minimize Eq.~\ref{eq:loss} with L-BFGS over the phase patterns $\{\varphi_j\}$, using the coupled optical and chemical forward model. Note that in our simulations we operate with normalized (dimensionless) inhibitor concentrations; consequently, the thresholds (e.g., $T_P$, $T_{\text{OP}}$, $T_I$) are selected empirically to yield stable printing outcomes. This is common practice in VAM/TVAM-style inverse design and avoids requiring absolute calibration of absorbed energy (laser power) and all reaction-rate constants.

\section{Results and discussion}
\label{sec:results}

In this section, we show that inhibitor diffusion degrades print fidelity for small features, and that this degradation can be partially mitigated by chemistry-aware hologram optimization. We further demonstrate that adding TEMPO, a less diffusive inhibitor, improves print quality (at the cost of increased required dose), and we validate these effects experimentally in 2D and 3D.

To facilitate direct comparison across all experiments and simulations, we use a fixed optical configuration and a consistent set of reconstruction and optimization parameters summarized in \autoref{tab:notations}. Unless otherwise stated, all results in this section use these settings.

\begin{table}[h]
\caption{Simulation and experimental parameters used throughout Sec.~\ref{sec:results}.}
\label{tab:notations}
\begin{tabularx}{\linewidth}{@{}lX@{}}
\toprule
Fourier lens focal length $f_\text{FL}$ & \SI{200}{\milli\meter} \\
Tube lens focal length $f_\text{TL}$ & \SI{100}{\milli\meter} \\
Objective lens focal length $f_\text{OL}$ & \SI{20}{\milli\meter} \\
Wavelength $\lambda$ & \SI{405}{\nano\meter} \\
SLM pixel pitch $\Delta x$ & \SI{8}{\micro\meter} \\
SLM resolution (used) $N \times N$ & $1200 \times 1200$ \\
Effective numerical aperture $\mathrm{NA}=\frac{N\,\Delta x\,f_\text{TL}}{2\,f_\text{FL}\,f_\text{OL}}$ & $0.12$ \\
Printing-plane pixel size $\Delta x_P$ & \SI{1.6875}{\micro\meter} \\
Printing-plane crop (off-axis) & $500\,\text{px} \times 500\,\text{px}$ ($\SI{844}{\micro\meter}\times\SI{844}{\micro\meter}$) \\
Optimization method / iterations & L-BFGS / $100$--$400$ \\
\added{Optimization time} & \added{\SI{30}{\second} to \SI{10}{\minute}}\\
Resin refractive index $n$ & $1.4803$ \\
Absorption coefficient $\mu$ & \SI{0.062}{\per\milli\meter} \\
Total exposure time $T_\text{exp}$ & \SI{8}{\second} \\
Number of time-multiplexed holograms $K$ & $20$ \\
Number of axial planes (simulation) & $10$--$30$ \\
Axial extent (simulation/print) & \SIrange{2}{3}{\milli\meter} \\
Laser power (experiment) & \SIrange{80}{400}{\micro\watt} \\
\bottomrule
\end{tabularx}
\end{table}

Based on the above configuration, the diffraction-limited lateral resolution (Abbe limit) is approximately \SI{1.7}{\micro\meter}. The corresponding axial extent of a diffraction-limited focus is on the order of \SI{50}{\micro\meter} under our effective NA; for extended objects, axial resolution is further degraded by the missing-cone limitation inherent to single-axis projection (\autoref{sec:missingcone}).

For each set of prints, laser power is empirically tuned within \SIrange{80}{400}{\micro\watt} to maximize print fidelity. Such calibration is standard in VAM, since the effective polymerization threshold depends on a nonlinear interplay between optical dose and the initial chemical state of the resin. Unless otherwise stated, 2D images are acquired through the same objective used for printing, using oblique illumination from below to enhance contrast.

\subsection{Oxygen diffusion correction}
\label{sec:oxygen_diffusion_correction}

    We evaluate how oxygen transport affects print fidelity by comparing two optimization settings: (i) an optical-only model that neglects oxygen diffusion and (ii) a diffusion-aware model that explicitly accounts for oxygen depletion and diffusion during exposure. Because radicals are generated with a spatially varying 3D distribution, oxygen is depleted non-uniformly, creating concentration gradients that drive diffusive replenishment and can blur fine features.
    
    \autoref{fig:diffusion_impact} summarizes the resulting forward-model predictions. When diffusion is ignored (top row), the optimization produces illumination patterns that deplete oxygen almost everywhere, including in intended void regions, which reduces inhibition and causes unintended polymerization. In contrast, diffusion-aware optimization (bottom row) reshapes the projected intensity to maintain higher inhibitor levels in void regions while still depleting oxygen within the target geometry, improving the separation between object and void distributions and yielding a more faithful thresholded reconstruction. 
    
    Overall, incorporating oxygen diffusion into the inverse design mitigates diffusion-induced artifacts which is consistent with prior observations in TVAM \cite{Orth_Webber_Zhang_Sampson_de}, although very fine features remain sensitive to inhibitor transport. This motivates using less diffusive inhibitors (e.g., TEMPO) and/or shorter exposure times to reduce the effective diffusion length $\ell \propto \sqrt{D\,T}$. \added{For further discussion see \autoref{sec:oxygen_variation}.}

    \subsection{Experimental diffusion coefficient measurement}
        \label{sec:experimental_diffusion_oxygen}
        The diffusion kernel in the forward model depends on material and environmental parameters (e.g., resin viscosity, temperature, and composition). We therefore require an experimental calibration of the oxygen diffusion coefficient for the specific resin used in this work.
        \begin{figure}[h]
            \centering
            \includegraphics[width=.5\textwidth]{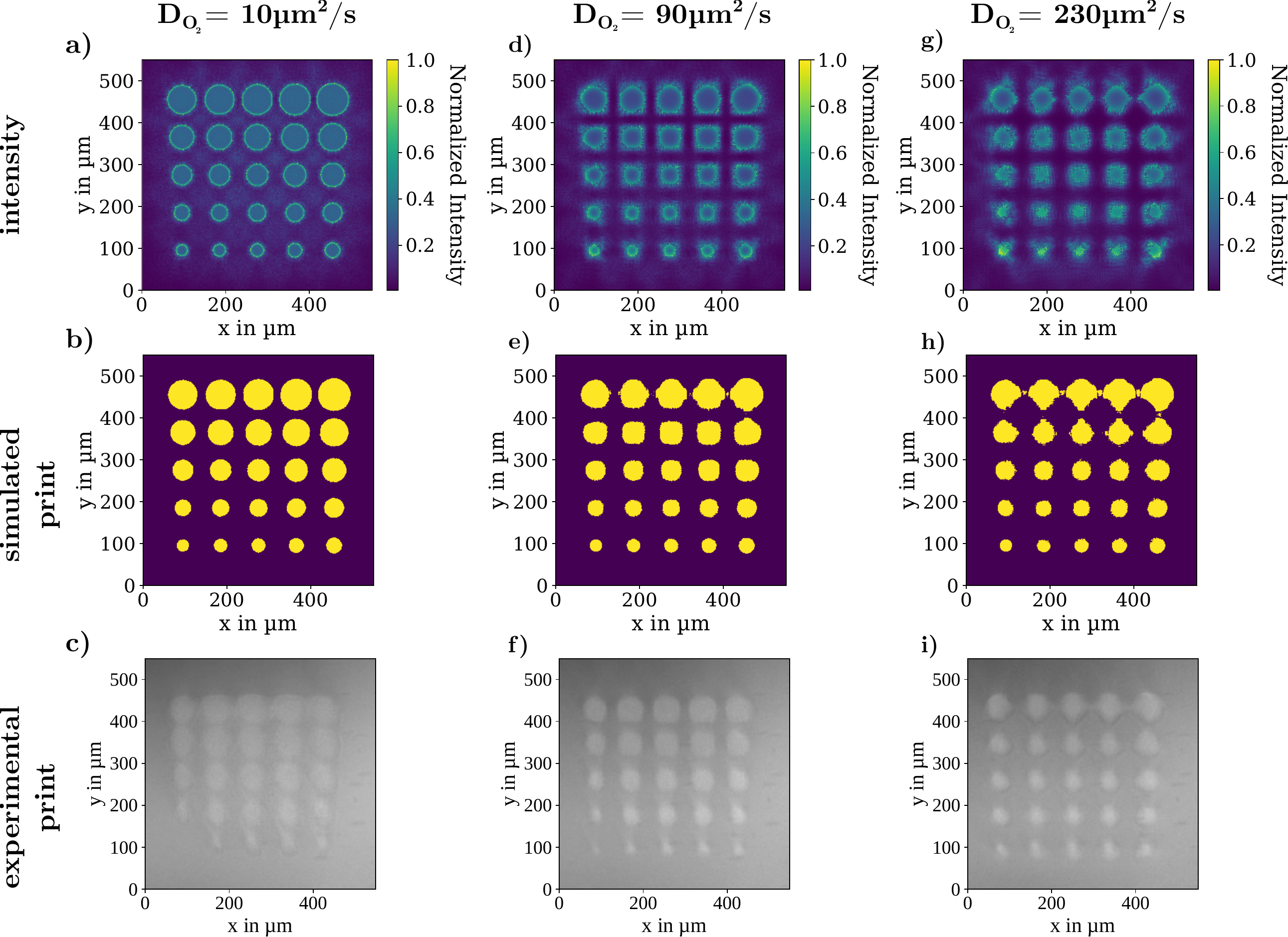}
            \caption{Experimental calibration of the oxygen diffusion coefficient. \textbf{First row)} Example optimized projected intensity for a candidate diffusion coefficient. \textbf{Second row)} Corresponding simulated print outcome. \textbf{Third row)} Experimental print imaged from above in a widefield configuration.}
            \label{fig:oxygen_diffusion}
        \end{figure}
        To estimate $D_{\ce{O2}}$, we fabricate a set of 2D test structure consisting of 25 cylindrical pillars with varying diameters (\autoref{fig:oxygen_diffusion}). 
        The diameters are ranging from \SI{30}{\micro\meter} to \SI{80}{\micro\meter} in steps of \SI{2}{\micro\meter}.
        For a range of candidate diffusion coefficients, we (i) optimize the projected illumination patterns using the diffusion-aware simulation model and (ii) project the resulting patterns into the prepared resin under identical experimental conditions.
        \autoref{fig:oxygen_diffusion}a) shows an example of an optimized intensity distribution for an assumed diffusion coefficient of $D=\SI{10}{\micro\meter\squared\per\second}$, which is predicted in simulation to yield the printed structure shown in \autoref{fig:oxygen_diffusion}b). 

    Experimentally, however, the corresponding projection produces the print shown in \autoref{fig:oxygen_diffusion}c), acquired with a widefield camera imaging the resin container from above. The observed mismatch between simulation and experiment indicates that $D=\SI{10}{\micro\meter\squared\per\second}$ underestimates oxygen transport in our resin.
        Notably, the smallest pillar is entirely missing. For $D_{\ce{O2}}=\SI{90}{\micro\meter\squared\per\second}$ we can see, that the smallest pillars are still underpolymerized. Small features are especially affected by diffusion and hence are better suited to estimate a rough value of the diffusion coefficient.

        We repeat this procedure for diffusion coefficients in the range $D_{\ce{O2}}=\SI{90}{\micro\meter\squared\per\second}$ to $D_{\ce{O2}}=\SI{280}{\micro\meter\squared\per\second}$. A best-match estimate is obtained by visually comparing experimental prints to their simulated counterparts across the full pillar set. Using this comparison, we find that $D_{\ce{O2}}=\SI{230\pm50}{\micro\meter\squared\per\second}$ provides the most consistent agreement. Smaller diffusion coefficients (e.g., $D_{\ce{O2}}=\SI{90}{\micro\meter\squared\per\second}$) tend to under-polymerize the smallest pillars, whereas larger values overcompensate diffusion and degrade feature fidelity. Because the present matching criterion is based on visual assessment rather than an automated metric, we report a conservative uncertainty of $\pm\SI{50}{\micro\meter\squared\per\second}$.
        This value is also in agreement with a quantitative experimental measurement in \autoref{app:oxygendiameter} of different pillar diameters.

        More accurate chemical diffusion measurements also require precise optical aberration correction as otherwise it is unclear whether the dominating error comes from a mismatch in the diffusion coefficient or whether it is aberration induced.

\subsection{TEMPO as additional inhibitor}
    Previous results in TVAM have demonstrated that adding TEMPO as a secondary inhibitor significantly improves printability \cite{Toombs_Luitz_Cook_Jenne_Li_Rapp, THIJSSEN2024106096}. Qualitatively, this benefit is attributed to the larger molecular size of TEMPO compared to oxygen, which results in lower diffusivity and helps preserve sharp polymerization boundaries.
    
    In this section, we incorporate quantitative modeling of TEMPO into the SHVAM inverse design framework and validate its impact through simulation and experiment. We utilize the resin formulation described in \autoref{sec:resinprep}, with relative concentrations of oxygen and TEMPO determined experimentally as detailed in \autoref{sec:concentrationscharac}.
    
\begin{figure}[h]
        \centering
        \includegraphics[width=1\linewidth]{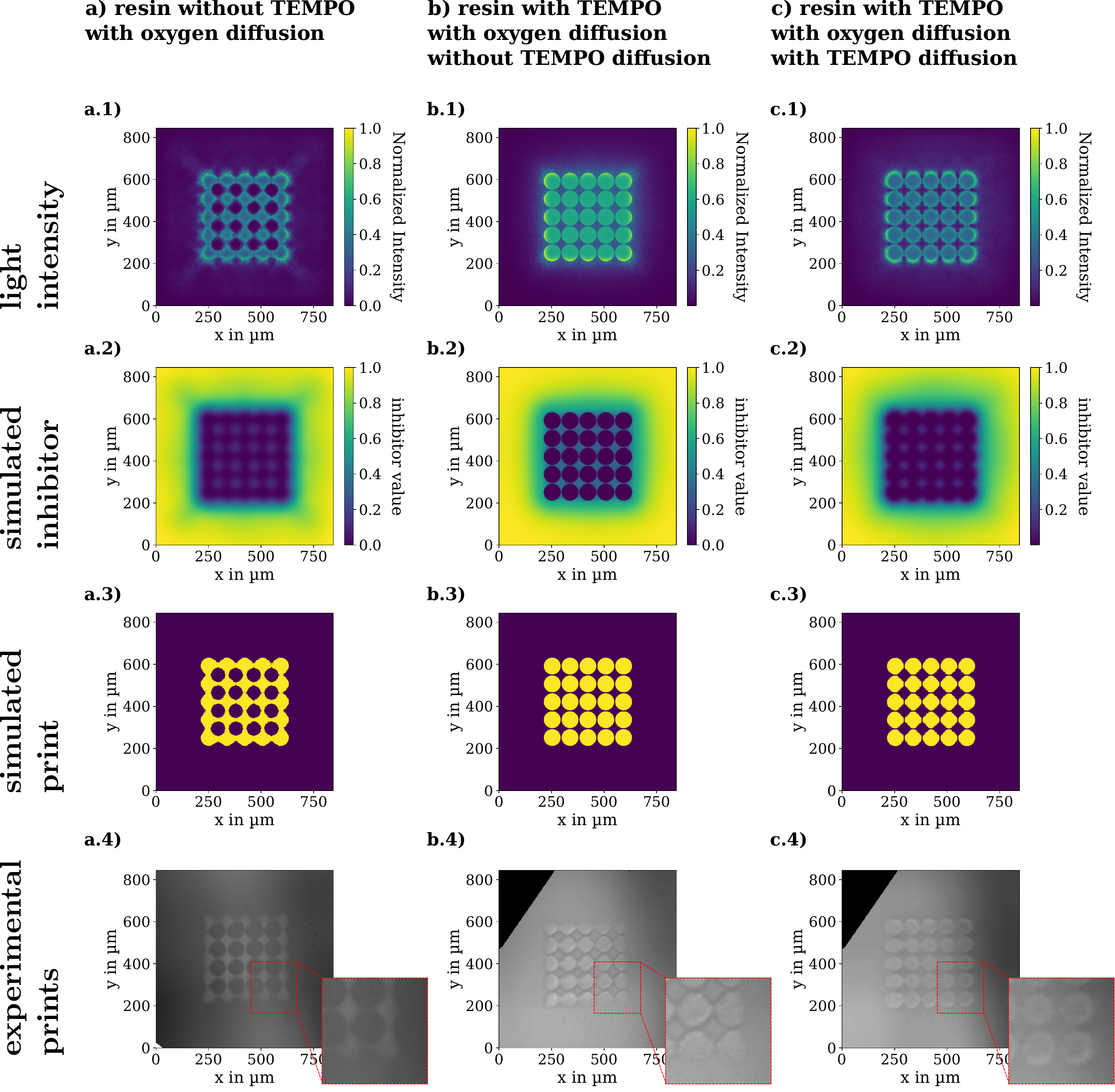}
        \caption{Lateral resolution test comparing different inhibitor configurations: \textbf{a)} oxygen only (with diffusion correction), \textbf{b)} oxygen and TEMPO (without TEMPO diffusion modeling), and \textbf{c)} oxygen and TEMPO (with optimized diffusion coefficients). Each row displays: \textbf{1)} projected light intensity, \textbf{2)} final inhibitor concentration, \textbf{3)} simulated print outcome, and \textbf{4)} experimental result.}
        \label{fig:resolutiontest}
    \end{figure}

    To assess the impact of inhibitor diffusion on lateral resolution, we designed a stress test consisting of a dense array of closely spaced pillars (\SI{85}{\micro\meter spacing}, \SI{80}{\micro\meter} diameter). The results are summarized in \autoref{fig:resolutiontest}.
    
    First, we examine a resin containing only oxygen, optimized with oxygen diffusion correction (\autoref{fig:resolutiontest}a). Although the projected intensity (\autoref{fig:resolutiontest}a.1) attempts to compensate for blur, the high diffusivity of oxygen results in a final inhibitor concentration field with poor contrast (\autoref{fig:resolutiontest}a.2). Consequently, the inhibitor is depleted across the inter-pillar gaps, leading to merged features in both the simulation (\autoref{fig:resolutiontest}a.3) and the experiment (\autoref{fig:resolutiontest}a.4). This confirms that even with algorithmic correction, oxygen diffusion limits the achievable pitch of dense structures.
    
    Next, we introduce TEMPO into the resin but neglect its diffusion in the optimization model (\autoref{fig:resolutiontest}b). The experimental print (\autoref{fig:resolutiontest}b.4) shows a marked improvement in structure definition compared to the oxygen-only case, confirming the chemical benefit of TEMPO. However, because the optimization assumes TEMPO is stationary, it fails to account for the slight diffusion that does occur. As a result, the outer pillars are under-polymerized and appear incomplete or missing in the experimental result.
    
    Finally, we optimize the patterns using the full coupled model with experimentally tuned diffusion coefficients of $D_{\ce{O2}}=\SI{230}{\micro\meter\squared\per\second}$ and $D_{\ce{TEMPO}}=\SI{110}{\micro\meter\squared\per\second}$ (\autoref{fig:resolutiontest}c). By explicitly accounting for the transport of both inhibitors, the optimizer boosts the dose at the boundaries to counteract dilution. The resulting experimental print (\autoref{fig:resolutiontest}c.4) shows the best agreement with the target design, with the boundary pillars fully preserved.
    
    While eliminating oxygen entirely to rely solely on TEMPO would theoretically yield superior results, preparing oxygen-free resins requires inert gas environments or vacuum processing, which significantly increases experimental complexity. Also, in cell-laden bioresins, adequate oxygen availability is required to prevent hypoxia.
    
    These results highlight that while diffusion-aware optimization can extend the resolution limits of VAM, fundamental physical constraints remain. Reducing the effective diffusion length would require, for example, resins with significantly higher viscosity (lowering $D$) or faster printing systems (lowering $T_\text{print}$). However, high viscosity complicates handling, and since the diffusion length scales as $\ell \propto \sqrt{D \cdot T_\text{print}}$, order-of-magnitude improvements in resolution will likely require substantial reductions in either the diffusion coefficient or the exposure time. Alternatively, more complex photochemical reaction mechanisms could be designed to mitigate these limitations.

\subsection{3D prints}
As final experiments, we demonstrate simple 3D prints. 

Due to the comparatively low numerical aperture of our projection optics (NA = 0.12), the axial confinement of the optical field is limited, resulting in poor axial resolution (i.e., a large depth-of-focus). Consequently, we restrict ourselves to objects without strong fine-scale variations along the optical axis.
For a theoretical discussion on printing with high NA systems, see \autoref{app:highna}.

\begin{figure}[h]
    \centering
    \includegraphics[width=1\linewidth]{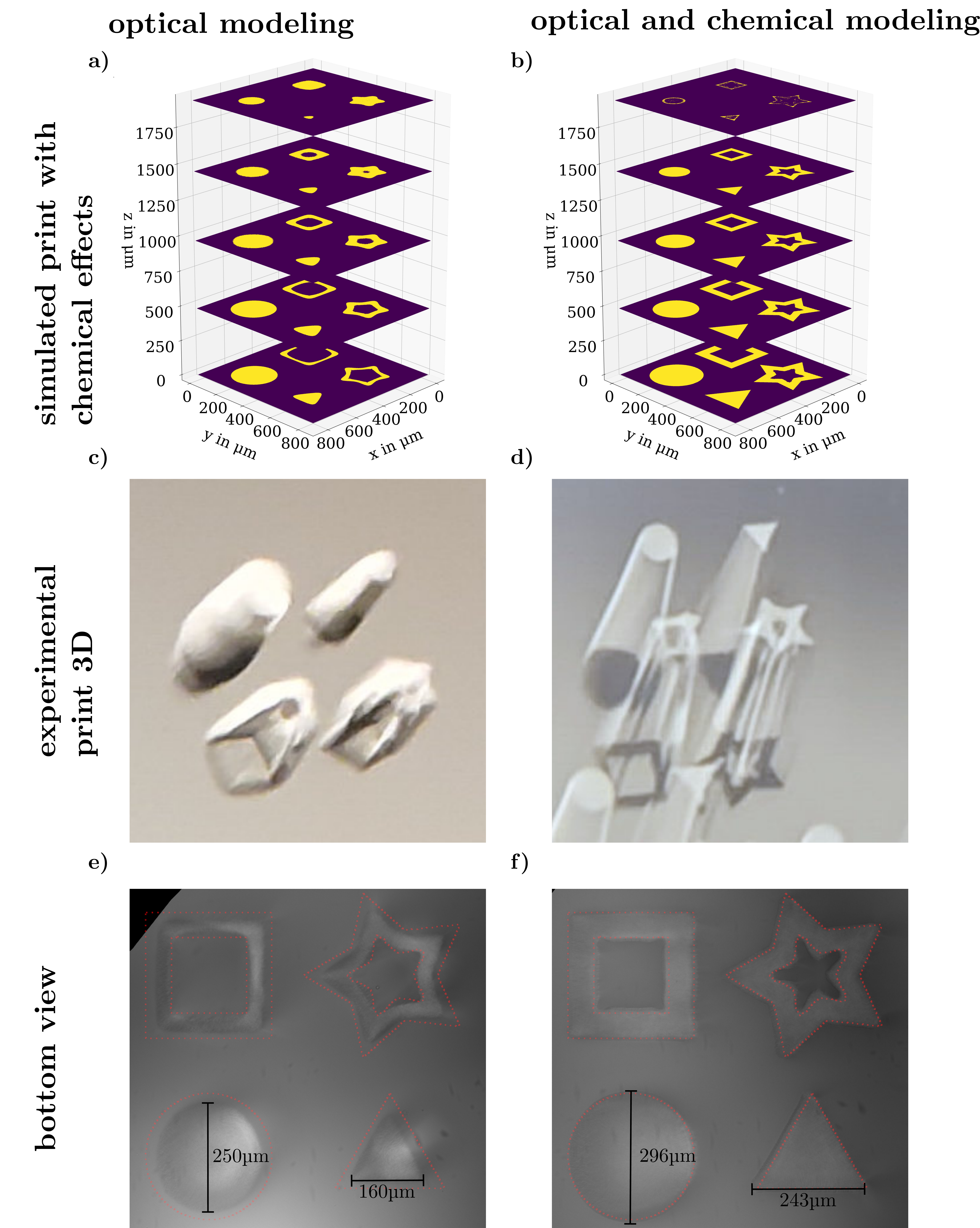}
    \caption{Comparison of 3D conical structures fabricated using optical-only optimization (left column) versus our diffusion-aware approach (right column). 
    \textbf{a, b)} Simulated print predictions evaluated using the full chemical forward model for \textbf{a)} optical-only patterns and \textbf{b)} chemically-optimized patterns.
    \textbf{c, d)} Experimental 3D views of the printed structures.
    \textbf{e, f)} Experimental 2D cross-sectional microscopy images (bottom view). Red dashed lines indicate the target design.
    The optical-only optimization \textbf{c, e)} results in feature shrinkage and occlusion of the central hollow channels (triangle side length: \SI{160}{\micro\meter}). In contrast, the chemically coupled optimization \textbf{d, f)} successfully compensates for diffusion, achieving better dimensions (triangle side length: \SI{243}{\micro\meter}) and preserving the hollow core structure.}
    \label{fig:starcone}
\end{figure}

\paragraph{Hollow, conical shaped structures}
As a first validation, we fabricated a set of hollow and solid conical structures aligned along the optical axis. The results are presented in \autoref{fig:starcone}. We compare two optimization strategies: one based on a purely wave-optical model (neglecting chemistry, left column) and one incorporating our coupled diffusion model (right column). To provide a fair comparison, the predicted outcomes for both strategies shown in \autoref{fig:starcone}a and \autoref{fig:starcone}b are simulated using the full optical-chemical forward model.

While patterns optimized solely with the wave-optical model reproduce the gross geometry, they suffer from reduced fidelity; specifically, internal hollow channels are partially occluded and sharp corners appear rounded (see \autoref{fig:starcone}c,d). Quantitative measurements in \autoref{fig:starcone}e,f confirm this loss of resolution: the experimental side length of the triangle was \SI{160\pm10}{\micro\meter} against a target of \SI{245}{\micro\meter}, while the cone diameter was \SI{250\pm10}{\micro\meter} against a target of \SI{280}{\micro\meter}. These experimental deviations are consistent with the outcome predicted by the chemical simulation in \autoref{fig:starcone}a.

In contrast, when diffusion effects are explicitly included in the inverse design, the hollow cores within the square- and star-shaped cones are preserved (see \autoref{fig:starcone}d,f), yielding prints that closely match the intended topology. This improvement is confirmed quantitatively, with the triangle side length and cone diameter measured at \SI{243\pm10}{\micro\meter} and \SI{296\pm10}{\micro\meter}, respectively.

Note that for all experimental prints, the global laser power was finely tuned to achieve the optimal trade-off between solidification and feature resolution.

\begin{figure}[h]
    \centering
    \includegraphics[width=1\linewidth]{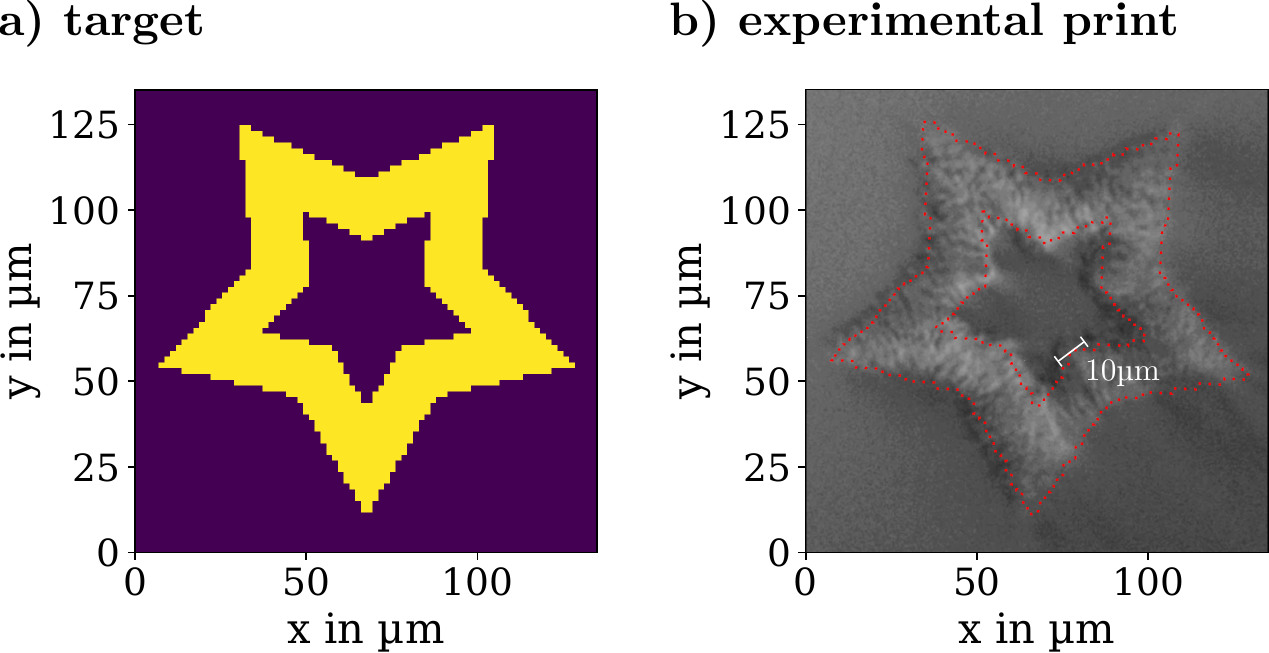}
    \caption{High-resolution printing demonstration. \textbf{a)} 2D cross section of the target hollow star geometry. \textbf{b)} Optical microscopy image of the printed structure, fabricated in \SI{1.5}{\second}. The system resolves lateral features of approximately \SI{10}{\micro\meter}, though some surface granularity remains due to optical speckle.}
    \label{fig:small_print}
\end{figure}

\paragraph{High-resolution print}
To probe the resolution limits of our system, we fabricated a hollow, conical star structure, shown in \autoref{fig:small_print}. The target geometry (\autoref{fig:small_print}a) contains fine features designed to test lateral confinement. The corresponding experimental result (\autoref{fig:small_print}b) demonstrates that SHVAM can resolve features on the order of \SI{10}{\micro\meter} with a total print time of just \SI{1.5}{\second}.

The printed structure exhibits a granular texture, which we attribute to speckle noise in the time-integrated dose distribution. Notably, our simulations based on the same optical model predict that even finer features are attainable (see \autoref{app:resolution}). This suggests that the current experimental resolution is limited primarily by projection quality (specifically speckle contrast and system aberrations) rather than by the inverse design formulation itself. Future improvements in holographic calibration (e.g., \cite{10.1145/3414685.3417802}) could reduce these artifacts and allow the system to approach the theoretical resolution limits predicted by simulation.

\begin{figure}[h]
    \centering
    \includegraphics[width=0.93\linewidth]{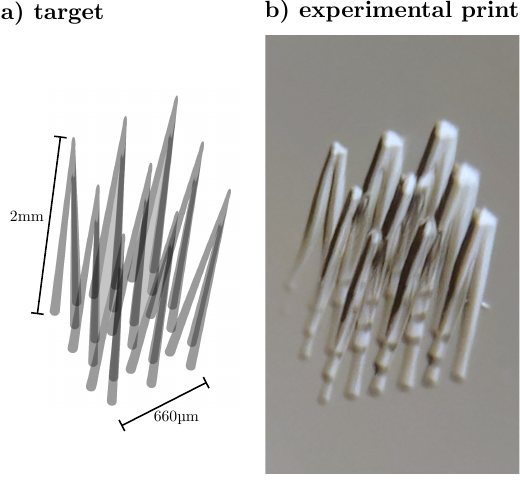}
    \caption{High-throughput fabrication demonstration. 
    A $3 \times 3$ array of hollow pyramids printed simultaneously in \SI{8}{\second}. 
    Despite the axial resolution limits imposed by the low-NA optics (visible as blunted tips), the system achieves parallel fabrication of multiple complex micro-structures across the full \SI{800}{\micro\meter} field of view. \textbf{a)} shows the target geometry. \textbf{b)} displays the final print.}
    \label{fig:pyramid_grid}
\end{figure}

\paragraph{High-throughput printing}
Finally, we demonstrate the high-through-put capabilities of SHVAM. \autoref{fig:pyramid_grid} displays a $3 \times 3$ grid of three-legged structures fabricated simultaneously in a single exposure sequence. While the low numerical aperture of our current objective limits the axial sharpness at the tips, the system successfully produces the entire array with consistent geometry. The total print time for this array was \SI{8}{\second}.

\paragraph{Printing speed \added{and Scalability}}
Based on our experimentally demonstrated field of view of \SI{800}{\micro\meter} (with $\sim\SI{10}{\micro\meter}$ lateral resolution) and a printable depth of \SI{3}{\milli\meter} (with $\sim\SI{150}{\micro\meter}$ axial resolution), we estimate a total addressable volume of approximately $~128\,000$ effective voxels. Given the print duration, this corresponds to a continuous production rate of approximately $16\,000$ voxels/s.
However, by using piston based phase light modulators (PLMs) or digital micromirror devices (DMDs) with Lee hologram projections \cite{holotvam}, pattern display rates can be increased from the current \SI{40}{\hertz} to $\gg \SI{1000}{\hertz}$. Additionally, by also increasing the laser power from around \SI{400}{\micro\watt} (this work) to \si{\milli\watt} regimes, printing rates of $> 1\,000\,000$ voxels/s could be achieved.
\added{Additionally to high voxel printing speeds, SHVAM is inherently scalable. Printing from one side allows easy implementation of large area scanning or mass manufacturing on conveyor belts}. \added{Another advantage of single-view projection lies in its ability to operate in mechanically or biologically constrained environments where multi-view access is not feasible. \cite{Urciuolo_Poli_Brandolino_Raffa_Scattolini_Laterza_Giobbe_Zambaiti_Selmin_Magnussen_et}.}

\paragraph{Axial resolution limitation}
\added{The axial resolution of a single-view system remains limited. In particular, the low-NA implementation considered here exhibits poor axial features. This is a well-known limitation in widefield microscopy (see \autoref{sec:missingcone}) and can only be mitigated by increasing the NA, at the expense of lateral field of view. However, simulations of high-NA configurations show substantially improved axial fidelity (see \autoref{app:highna}). Such systems, however, require more extensive calibration, particularly along the vertical axis.}

\added{As a rule of thumb, the best-case lateral resolution, ignoring chemical effects, is given by $\Delta x = \frac{\lambda}{2\mathrm{NA}}$, while the axial resolution is approximately $\Delta z \approx \frac{2\lambda}{\mathrm{NA}^2}$. However, the axial resolution is not isotropic and depends on the lateral feature content, as illustrated by the frequency support in \autoref{fig:missingcone}.}

\section{Conclusion}
In this work, we introduced \emph{Single-View Holographic Volumetric Additive Manufacturing} (SHVAM), a computational fabrication framework that synthesizes 3D micro-structures from a single optical axis without moving parts. By shifting the complexity of volumetric shaping from mechanical hardware to algorithmic design, SHVAM utilizes time-multiplexed phase holography to sculpt time-integrated (effectively incoherent) dose distributions within a photosensitive resin. Central to our approach is a coupled differentiable forward model that integrates wave optics with a simplified inhibition--diffusion photochemistry surrogate.

Our experiments demonstrate that at the micro-scale, optical optimization alone is insufficient: transport of inhibitors (oxygen and TEMPO) significantly alters the effective printed geometry. By explicitly modeling these chemical dynamics within the inverse design loop, we can pre-compensate diffusion-driven blur and improve print fidelity. We validate this approach by fabricating 3D geometries with lateral features down to \SI{10}{\micro\meter} in exposure times as short as \SI{1.5}{\second}. Moreover, higher-refresh-rate modulators (e.g., piston-based phase light modulators) could enable sub-second exposures, increasing throughput while reducing diffusion-induced degradation by shortening the relevant transport time.

Our current prototype highlights key trade-offs inherent to single-view projection, including limited axial resolutions due to the low effective NA as well as artifacts from coherent projection (speckle and aberrations). While speckle and aberrations can be mitigated through improved holographic calibration and system characterization, improving axial resolution in a single-view architecture will require higher-NA optics (with the corresponding reduction in field of view, see \autoref{app:highna}) and/or additional degrees of freedom.

More broadly, our results suggest that coupling wave-optical inverse design with inhibition and diffusion modeling provides a practical path toward higher-fidelity volumetric microfabrication. Crucially, the photochemical formulation presented here is agnostic to the optical delivery method and can be readily integrated into other volumetric printing approaches — such as TVAM — to push resolution beyond current optical-only optimization limits.

\begin{acks}
    We thank Jonathan Dong for valuable discussions on optical theory and Baptiste Nicolet for proofreading the manuscript.
    During the preparation of this work, the authors used generative AI models to enhance the quality of the work. In particular, AI models were used to write source code and to improve the grammar, spelling, and clarity of the manuscript’s text (including this sentence).
    This project has received funding from the Swiss National Science Foundation 2000-1-240074 under grant number 10007068 - Neural precision holographic volumetric additive manufacturing and from the Swiss National Science Foundation Return CH Post- doc.Mobility P5R5-3\_235066 (R.R.). Christophe Moser is a shareholder of Readily3D SA. In addition, Christophe Moser and Felix Wechsler are inventors on pending patent application WO2025223658A1.
\end{acks}

\bibliographystyle{ACM-Reference-Format}
\bibliography{references}

\appendix{
\setcounter{figure}{0}
\renewcommand{\thefigure}{A\arabic{figure}}

\section{Optical setup}
    \label{sec:opticalsetup}
    A \SI{405}{\nano\meter} laser (Coherent OBIS LX SF 405-40, narrow-linewidth, single-longitudinal-mode) was used as the light source. The beam was collimated onto the spatial light modulator (SLM). A half-wave plate was used to align the beam polarization with the SLM. 
    
    The SLM was a Meadowlark E-19x12-400-800-HDM8 with a resolution of $1920\times1200$ pixels and a pixel pitch of \SI{8}{\micro\meter}. A Fourier lens with focal length $f_\text{FL}=\SI{200}{\milli\meter}$ (ACT508-200-A-ML, Thorlabs) was used, and an off-axis diffraction order was selected using a circular aperture. A tube lens with focal length $f_\text{TL}=\SI{100}{\milli\meter}$ (AC508-100-A-ML, Thorlabs) relayed the beam to the objective.
    
    For focusing into the resin, we used an objective lens with focal length $f_\text{OL}=\SI{20}{\milli\meter}$ (M Plan APO 10x, Mitutoyo), with a theoretical numerical aperture of $\mathrm{NA}=0.28$. In our setup, the back aperture of the objective was intentionally underfilled, resulting in a lower effective NA to increase the field of view.
    
    For imaging, the same microscope objective was used in combination with a dichroic mirror to block the blue illumination light partially. The imaging arm employed a \SI{120}{\milli\meter} tube lens and a CMOS camera (Basler ace 2 a2A4504-18umPRO).

\section{Resin preparation}
    \label{sec:resinprep}
    We prepared the resin by mixing DUDMA (Sigma-Aldrich, USA) and PEGDA 700 (Sigma-Aldrich, USA) at a 4:1 weight ratio. As the photoinitiator, we used diphenyl(2,4,6-trimethylbenzoyl)phosphine oxide (TPO, \SI{97}{\percent}; Sigma-Aldrich), which was first dissolved in \SI{99}{\percent} Isopropyl alcohol (IPA) and then incorporated using a planetary mixer (KK-250SE, Kurabo). As an additional inhibitor, we used 2,2,6,6-tetramethylpiperidine 1-oxyl (TEMPO, \SI{98}{\percent}; Sigma-Aldrich); TEMPO was likewise dissolved in \SI{99}{\percent} IPA before being mixed into the resin.

    The resin had a measured refractive index of $n=1.4803\pm0.003$ and an absorption coefficient of $\mu = \SI{0.062}{\per\milli\meter}$. For sample preparation, small droplets of resin were dispensed onto a microscope slide and confined with a rubber spacer ring of thickness \SIrange{1.5}{3}{\milli\meter}. A cover slip was placed on top of the ring to seal the volume.
    The printed volume spanned the entire gap between the glass interfaces.

\section{Oxygen diffusion coefficient}
\label{app:oxygendiameter}

In addition to the qualitative (visual) estimate reported in \autoref{sec:experimental_diffusion_oxygen}, we perform a simple quantitative consistency check based on measured feature sizes. Specifically, we fabricate three isolated pillar targets with nominal diameters of \SI{20}{\micro\meter}, \SI{40}{\micro\meter}, and \SI{60}{\micro\meter}. For a range of candidate oxygen diffusion coefficients $D_{\ce{O2}}$, we re-optimize the hologram patterns and compare the resulting simulated pillar diameters to those measured experimentally.

As shown in \autoref{fig:oxygendiameter}, small diffusion coefficients ($D_{\ce{O2}} \lesssim \SI{100}{\micro\meter\squared\per\second}$) lead to severe under-polymerization, and the smallest pillars are partially formed or missing. In contrast, diffusion coefficients in the range \SIrange{150}{270}{\micro\meter\squared\per\second} yield the closest agreement between simulated and experimental diameters across the three pillar sizes.

\begin{figure}[t]
    \centering
    \includegraphics[width=.9\linewidth]{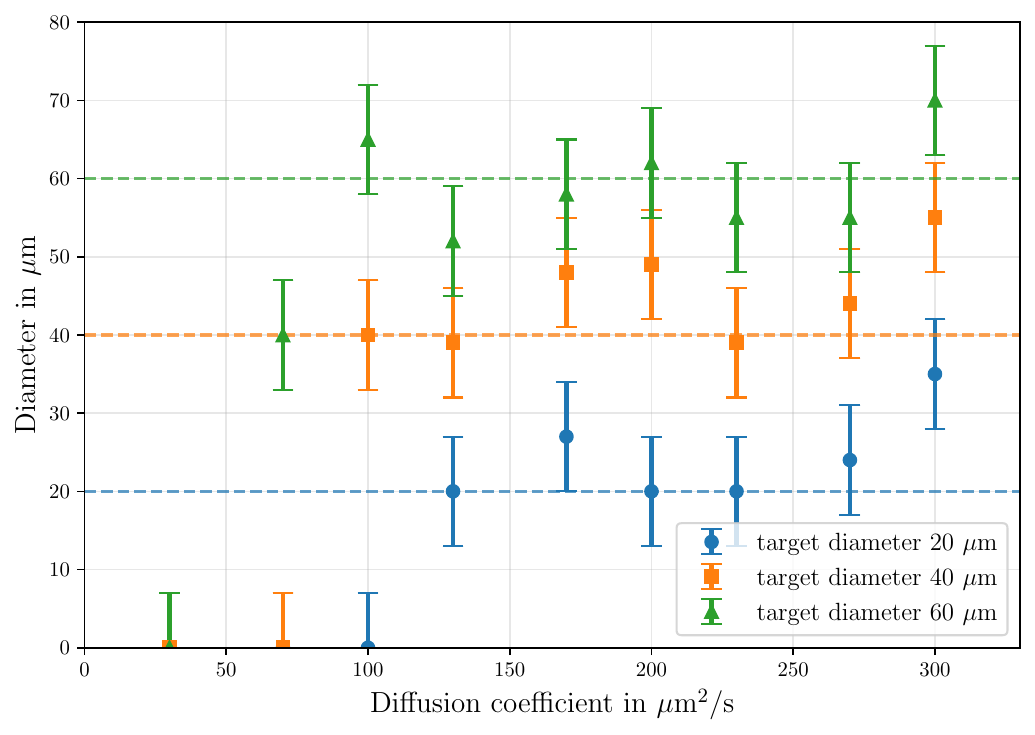}
    \caption{Measured (experiment) pillar diameters for different assumed oxygen diffusion coefficients $D_{\ce{O2}}$, evaluated on pillar targets with nominal diameters of \SI{20}{\micro\meter}, \SI{40}{\micro\meter}, and \SI{60}{\micro\meter}.}
    \label{fig:oxygendiameter}
\end{figure}

We note that diameter measurements are inherently noisy in our setup because the apparent pillar boundary depends on the chosen exposure and focus settings during imaging, introducing operator-dependent uncertainty. Given these limitations, we select $D_{\ce{O2}}=\SI{230}{\micro\meter\squared\per\second}$ as a representative value, consistent with the estimate reported earlier.

\section{Experimental characterization of relative inhibitor concentrations}
    \label{sec:concentrationscharac}
    The base resin contained a photoinitiator (TPO) concentration of $c_{\ce{TPO}}=\SI{1.2}{\milli\molar}$. This batch was then split into two portions. To the second portion, we added \SI{2.5}{\milli\gram} TEMPO to \SI{38.46}{\gram} resin, corresponding to $c_{\ce{TEMPO}}=\SI{0.46}{\milli\molar}$.

    \begin{figure}
        \centering
        \includegraphics[width=1\linewidth]{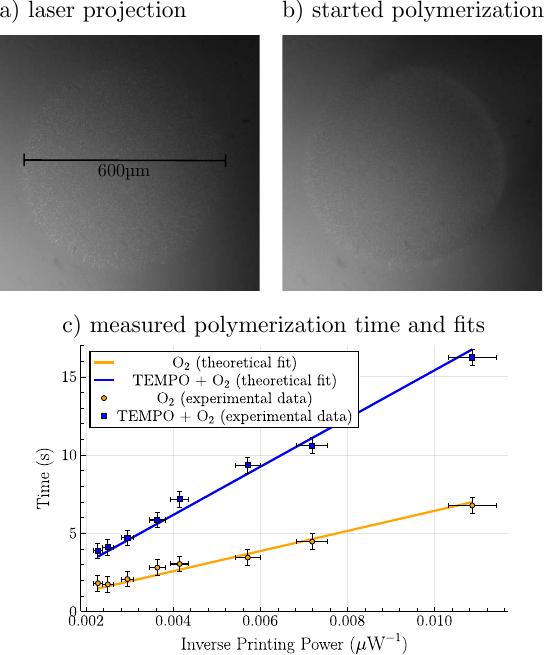}
        \caption{Inhibitor depletion-time measurements for two resins: one containing oxygen only and one containing oxygen and TEMPO. \textbf{a)} Projected circular disc pattern. \textbf{b)} Polymerized region after several seconds of illumination. \textbf{c)} Measured exposure time versus projected intensity, together with the fitted model.}
        \label{fig:oxygenconcentrationfit}
    \end{figure}
    
    While $c_{\ce{TEMPO}}$ is known from the preparation, the dissolved oxygen concentration in the resin is not. We therefore performed a calibration experiment to estimate $c_{\ce{O2}}$. Following recent work \cite{Zhang_DeHaan_Houlahan_Sampson_Webber_Orth_Lacelle_Gaburici_Lam_Deore_etal._2025}, we assume that the time prior observable polymerization is dominated by inhibitor depletion and is therefore approximately proportional to the inhibitor concentration. Concretely, we varied the projected intensity and measured the exposure time until polymerization became visible. \autoref{fig:oxygenconcentrationfit}a) illustrates the projected pattern, and \autoref{fig:oxygenconcentrationfit}b) shows the polymerized region after several seconds of illumination. A circular disc of \SI{600}{\micro\meter} diameter was projected. The diameter was chosen to be sufficiently large to neglect inhibitor diffusion during the exposure.
    
    Using the analytical model in \cite{Zhang_DeHaan_Houlahan_Sampson_Webber_Orth_Lacelle_Gaburici_Lam_Deore_etal._2025}, we fit the oxygen concentration $c_{\ce{O2}}$ and a correction factor $\eta$ (absorbing error in the power measurement and quantum efficiency of TPO) to the measured exposure times. The depletion time for oxygen is
    \begin{equation}
        T_{\ce{O2}} = \frac{N_A \cdot h \cdot \nu \cdot c_{\ce{O2}}}{2 \cdot \eta \cdot \varepsilon \cdot I \cdot c_{\ce{TPO}}},
        \label{eq:depletion_time_ox}
    \end{equation}
    and for TEMPO
    \begin{equation}
        T_{\ce{TEMPO}} = \frac{N_A \cdot h \cdot \nu \cdot c_{\ce{TEMPO}}}{2 \cdot \eta \cdot \varepsilon \cdot I \cdot c_{\ce{TPO}}},
        \label{eq:depletion_time_tempo}
    \end{equation}
    where $N_A$ is the Avogadro constant, $h$ is the Planck constant, $\nu=\SI{740}{\tera\hertz}$ is the laser frequency.  $c_{\ce{TPO}}=\SI{1.2}{\milli\molar}$, and $c_{\ce{TEMPO}}=\SI{0.46}{\milli\molar}$ are the concentrations of TPO, and TEMPO, respectively. $\varepsilon=\SI{500}{\liter\per\centi\meter\per\mol}$ is the molar absorption coefficient (estimated from the TPO concentration and measured absorbance), and $I$ is the illumination intensity.
    
    We performed two sets of time measurements for different illumination intensities:
    \begin{itemize}
        \item Resin containing oxygen only (no added TEMPO), for which we approximate the polymerization time as $T_{\ce{O2}}$ (Eq.~\ref{eq:depletion_time_ox}).
        \item Resin containing oxygen and TEMPO, for which we approximate the polymerization time as $T_{\ce{O2}} + T_{\ce{TEMPO}}$ (Eqs.~\ref{eq:depletion_time_ox} and \ref{eq:depletion_time_tempo}).
    \end{itemize}
    Strictly, depletion time is not identical to the polymerization time; however, we use this approximation because inhibitor depletion dominates the observed delay before polymerization \cite{Zhang_DeHaan_Houlahan_Sampson_Webber_Orth_Lacelle_Gaburici_Lam_Deore_etal._2025}.
    The fit yields $c_{\ce{O2}}=\SI{0.33\pm0.03}{\milli\molar}$ and $\eta=0.36\pm0.02$, as shown in \autoref{fig:oxygenconcentrationfit}c).    
    In the simulations, we use relative (normalized) concentrations rather than absolute values. We therefore set $c[\ce{O2}]=1$ and $c[\ce{TEMPO}]=1.397$.

\section{Oxygen Diffusion Coefficient Variation}
\label{sec:oxygen_variation}
\added{To understand the impact of photochemical parameters on print quality, we systematically vary the oxygen diffusion coefficient and analyze the resulting print fidelity. \autoref{fig:oxygen_diffusion_variation} shows simulation results for different oxygen diffusion coefficients ranging from low to high values. The simulated printing time was $\SI{3}{\second}$}.

\added{For very low diffusion coefficients ($D_{O_2} = \SI{1}{\micro\meter^2/s}$), we observe near-perfect print results with loss values approaching zero, indicating convergence of our coupled optimization framework. However, as the oxygen diffusion coefficient increases, the joint wave-optical and photochemical optimization encounters fundamental resolution limits. At very high diffusion values ($D_{O_2} = \SI{1000}{\micro\meter^2/s}$), the target geometry becomes significantly distorted due to excessive chemical blur that cannot be compensated by optical pre-correction.}

\added{This demonstrates that while chemical correction substantially improves print fidelity compared to optical-only approaches, it is ultimately bounded by the underlying photochemical processes.
The characteristic diffusion length scales roughly with $\sqrt{D \cdot T_\text{print}}$, so reducing either the diffusion coefficient or the printing time reduces this limit. 
The target geometry used in this analysis has an axial extent of \SI{4}{\milli\meter}, representing a extended 3D structure rather than a simplified 2D case.}

\begin{figure}[h]
    \centering
    \includegraphics[width=1\linewidth]{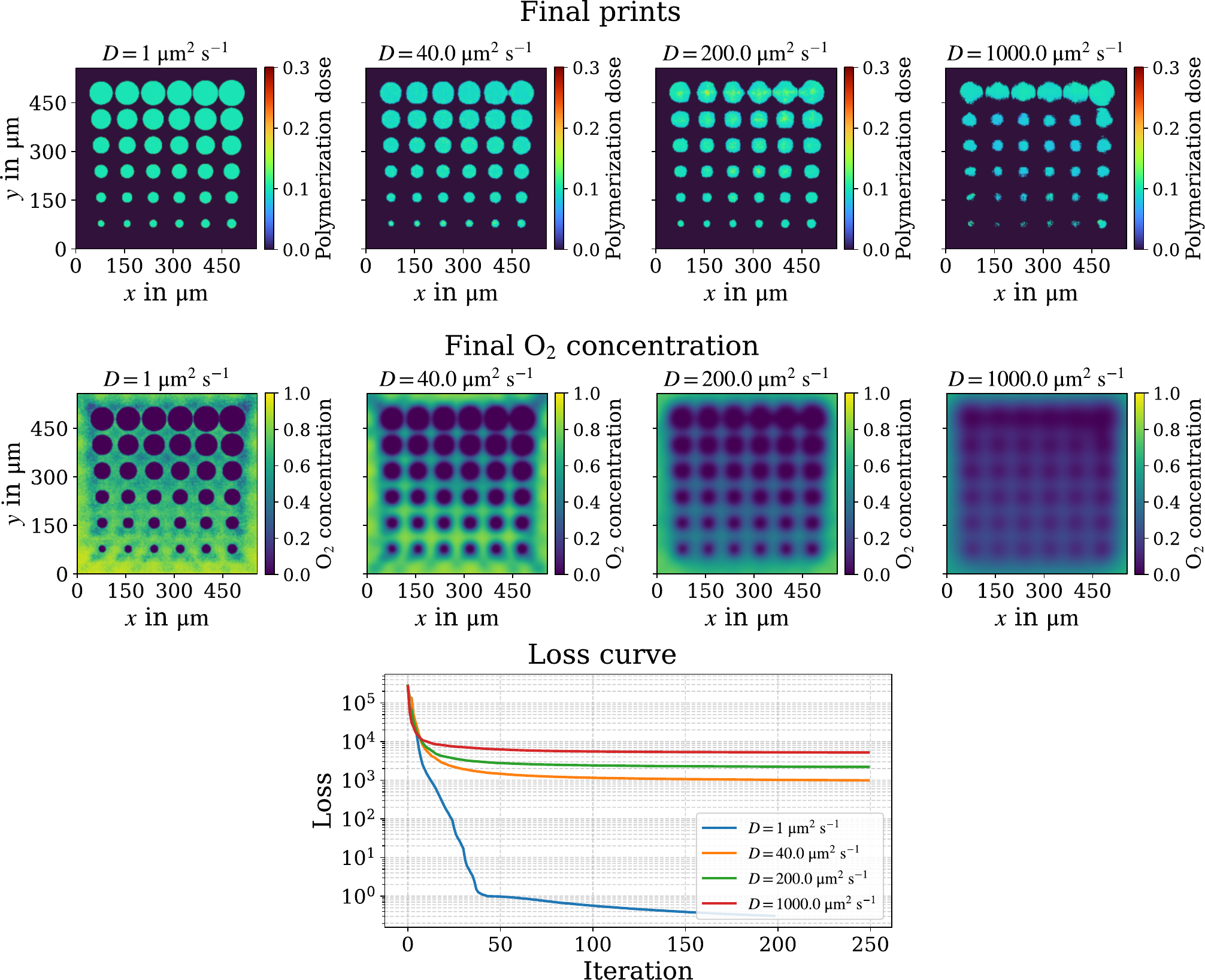}
    \caption{
    \added{
    Simulations showing the effect of oxygen diffusion coefficient on print quality and optimization convergence. Our coupled wave-optical and photochemical optimization framework mitigates chemical blur for moderate diffusion values but encounters fundamental limits at extreme values.
    \textbf{First row:} Final polymerized structures showing progressive degradation with increasing diffusion.
    \textbf{Second row:} Corresponding oxygen concentration distributions.
    \textbf{Third row:} Optimization loss convergence curves demonstrating the relationship between chemical parameters and achievable print fidelity.}}
    \label{fig:oxygen_diffusion_variation}
\end{figure}

 \section{Motivation to use several patterns}
\label{sec:multiple_patterns}
Instead of an electronically controllable SLM, one could also employ a static phase mask \cite{lin2026singleexposureholographic3dprinting}. A key advantage of a fabricated mask is that it can be made significantly larger than the active aperture and pixel count of a single SLM, thereby relaxing constraints on the addressable spatial bandwidth.

There are, however, two fundamental disadvantages of relying on a \emph{single} coherent phase pattern. First, a single coherent realization is more susceptible to experimentally induced coherent speckle. In contrast, displaying multiple distinct patterns and time-multiplexing them reduces speckle through temporal averaging, yielding a smoother deposited dose.

Second, and more importantly for volumetric patterning, a single 2D phase pattern cannot in general produce volumetric intensity distributions as rich as those achievable by a time-multiplexed set of patterns. The underlying reason is physical: for a monochromatic field, the complex electric field $E(\mathbf{r})$ in the volume is constrained by the Helmholtz equation,
\begin{equation}
\left(\nabla^2 + k^2\right) E(\mathbf{r}) = 0,
\end{equation}
(with $k = 2\pi/\lambda$ in the medium). Thus, a single phase pattern defines boundary conditions at the modulator plane and generates a \emph{single} coherent field $E(\mathbf{r})$ in the volume, from which the intensity follows as
\begin{equation}
I(\mathbf{r}) = |E(\mathbf{r})|^2.
\end{equation}
Consequently, the attainable intensity distributions are restricted to those that can arise as the squared magnitude of a Helmholtz-consistent field produced from that single boundary condition.

\begin{figure}[h]
    \centering \includegraphics[width=1\linewidth]{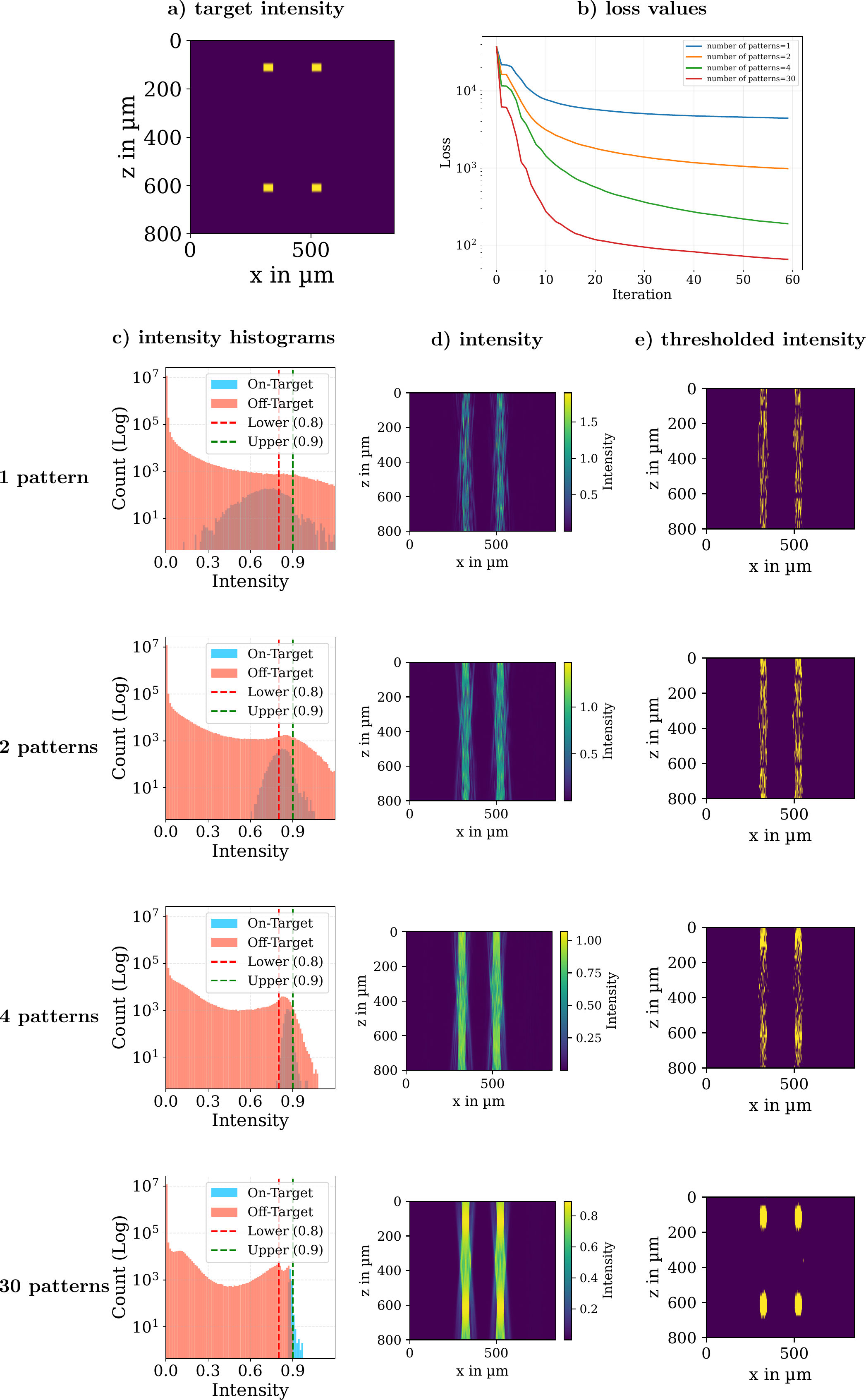}
    \caption{Simulation results of the intensity distribution for a different amount of phase patterns.
    \textbf{a)} Target. \textbf{b)} Loss curves for 1, 2, 4, and 30 phase patterns. \textbf{c)} Intensity histograms for different pattern counts. \textbf{d)} Summed intensities. \textbf{e)} Thresholded intensities at threshold 0.843.}
    \label{fig:onevsmultiple}
\end{figure}

By contrast, time-multiplexing $K$ phase patterns produces an \emph{incoherent} dose accumulation (since the resin integrates intensities),
\begin{equation}
I(\mathbf{r}) = \sum_{k=1}^{K} |E_k(\mathbf{r})|^2,
\end{equation}
where each field $E_k(\mathbf{r})$ individually satisfies the Helmholtz equation. Importantly, the \emph{sum of intensities} is not itself required to correspond to the intensity of any single Helmholtz solution, i.e., there does not generally exist an $E(\mathbf{r})$ such that $|E(\mathbf{r})|^2 = \sum_k |E_k(\mathbf{r})|^2$. This decouples the achievable dose distribution from the constraints imposed by any single coherent field realization and constitutes a fundamental physical motivation for using multiple patterns.

To demonstrate this effect, we optimize phase patterns such that their thresholded intensity matches the target distribution shown in \autoref{fig:onevsmultiple}a), consisting of four prisms of width \SI{40}{\micro\meter} and length \SI{30}{\micro\meter}, with a spacing of roughly \SI{500}{\micro\meter}.
As loss function for the optimization, we use a threshold-inspired penalty similar to those employed throughout this work:
\begin{equation}
    \mathcal{L} =
    \sum_{v \in \text{object}} 10 \cdot \left|\mathrm{ReLU}(0.9 - I_v)\right|^2
    +
    \sum_{v \notin \text{object}} \left|\mathrm{ReLU}(I_v - 0.8)\right|^2,
    \label{eq:loss2}
\end{equation}
where $I_v$ denotes the (time-multiplexed) intensity at voxel $v$.

Performing this optimization with an L-BFGS optimizer yields the loss curves shown in \autoref{fig:onevsmultiple}b).
The resulting intensity histograms are depicted in \autoref{fig:onevsmultiple}c), while \autoref{fig:onevsmultiple}d) shows the summed intensities and \autoref{fig:onevsmultiple}e) shows the corresponding thresholded intensities at threshold value 0.843.

We can clearly identify that a single phase pattern is not able to craft a 3D light intensity distribution in which laterally separated target objects remain well confined and separated in the volume. In practice, the field required to illuminate objects of this lateral extent tends to exhibit an extended axial depth-of-focus (analogous to the long Rayleigh range of a wide-waist Gaussian beam), which leads to unwanted axial spreading and overlap of the high-intensity regions.

As the number of time-multiplexed patterns increases, the projection becomes effectively more incoherent in the sense of dose accumulation: each pattern can contribute dose to only a subset of the target voxels, while depositing less dose elsewhere. Only their sum yields a sufficiently selective 3D intensity distribution such that the thresholded result matches the desired objects.
With rapidly refreshing displays such as piston based phase light modulators (PLM) (refresh rates $\gg \SI{1000}{\hertz}$) the spatial bandwidth can also be much further increased beyond the one of a single phase mask.

\section{Current resolution limits}
\label{app:resolution}
\autoref{fig:starcone_smaller} shows a challenging hollow conical target that probes the practical resolution limits of our current SHVAM prototype. In simulation, the coupled forward model predicts that the structure should be printable with the desired internal hollow star.

\begin{figure}[h]
    \centering
    \includegraphics[width=1\linewidth]{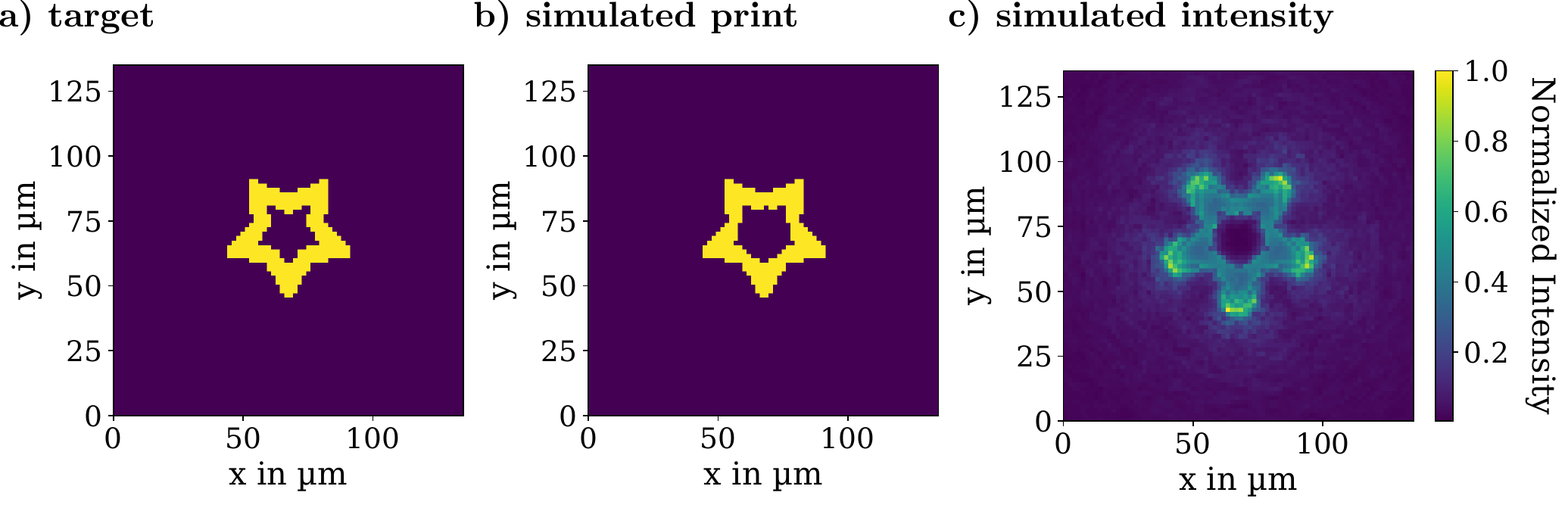}
    \caption{Resolution limits in simulation. \textbf{a)} Target geometry. \textbf{b)} Predicted print outcome (simulation). \textbf{c)} Corresponding optimized projection intensity.}
    \label{fig:starcone_smaller}
\end{figure}

Experimentally, however, we were not able to reliably preserve the full hole diameter, and the hollow core partially closes. We attribute this discrepancy primarily to residual coherent speckle and unmodeled system imperfections in the projection path, including SLM-related artifacts (e.g., phase nonidealities) and small optical misalignments or aberrations that are not captured in our idealized wave-optical model.

\section{Missing cone}
    \label{sec:missingcone}
    
    The limited axial resolution of a single-axis projection geometry can be understood from its spatial-frequency support, analogous to the optical transfer function (OTF) of incoherent widefield imaging \cite{Mertz_2019}. In this picture, the accessible 3D frequency region is given by the autocorrelation of the Ewald-sphere cap defined by the system numerical aperture (NA). Because lateral and axial frequencies scale differently, the resulting support is highly anisotropic and exhibits a characteristic \textit{missing cone} along the axial frequency axis.
    
    \autoref{fig:missingcone} visualizes this frequency support (values of the $|\text{OTF}| > 0$) for three numerical apertures: \textbf{a)} $\mathrm{NA}=0.12$, \textbf{b)} $\mathrm{NA}=0.36$, and \textbf{c)} $\mathrm{NA}=0.72$.
    
    \begin{figure}[h]
        \centering
        \includegraphics[width=0.8\linewidth]{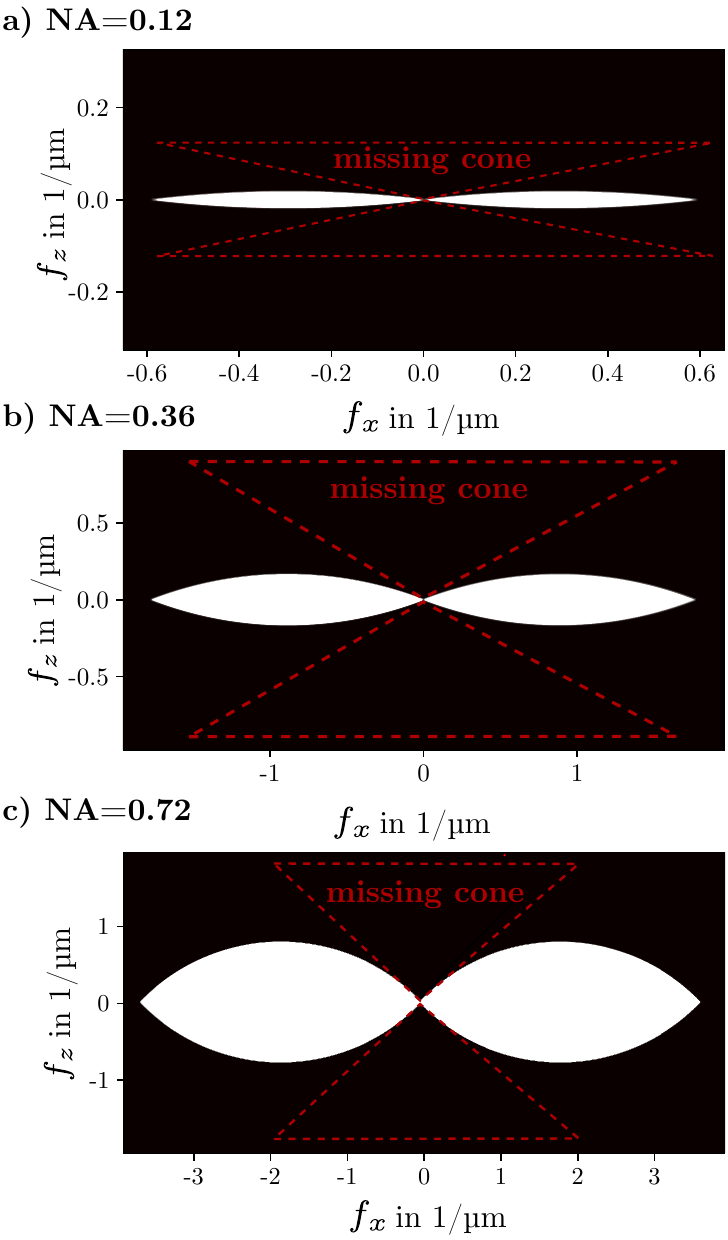}
        \caption{Spatial-frequency support (OTF) of a single-axis projection system for \textbf{a)} $\mathrm{NA}=0.12$, \textbf{b)} $\mathrm{NA}=0.36$, and \textbf{c)} $\mathrm{NA}=0.72$. Note the different scaling of lateral and axial frequency axes. The cone-shaped gap around the axial frequency axis corresponds to missing axial information and limits axial confinement.}
        \label{fig:missingcone}
    \end{figure}
    
    At $\mathrm{NA}=0.12$, axial frequency coverage is extremely limited, implying poor axial confinement and strong elongation of features along the optical axis. Increasing the NA to 0.36 substantially expands the accessible axial bandwidth, and $\mathrm{NA}=0.72$ provides good (though still anisotropic) coverage compared to lateral frequencies. In addition, at low lateral frequencies the axial coverage remains sparse, which particularly degrades the reproduction of objects that are thin or sharply varying along $z$.
    
    This missing-cone limitation is fundamental to single-view projection and cannot be removed by optimization alone; it requires additional angular diversity, for example by introducing a second projection arm to fill in the missing axial frequencies.

\section{High-NA resolution capabilities}
\label{app:highna}

To assess the theoretical volumetric resolution limits of SHVAM beyond our experimental prototype, we perform simulations using higher numerical aperture (NA) configurations. For this analysis, we focus solely on the optical constraints by disabling the inhibitor diffusion model, thereby isolating the diffraction-limited performance. We select the community-standard \#3DBenchy model as a challenging volumetric test target.

\begin{figure*}[t]
    \centering
    \includegraphics[width=1\linewidth]{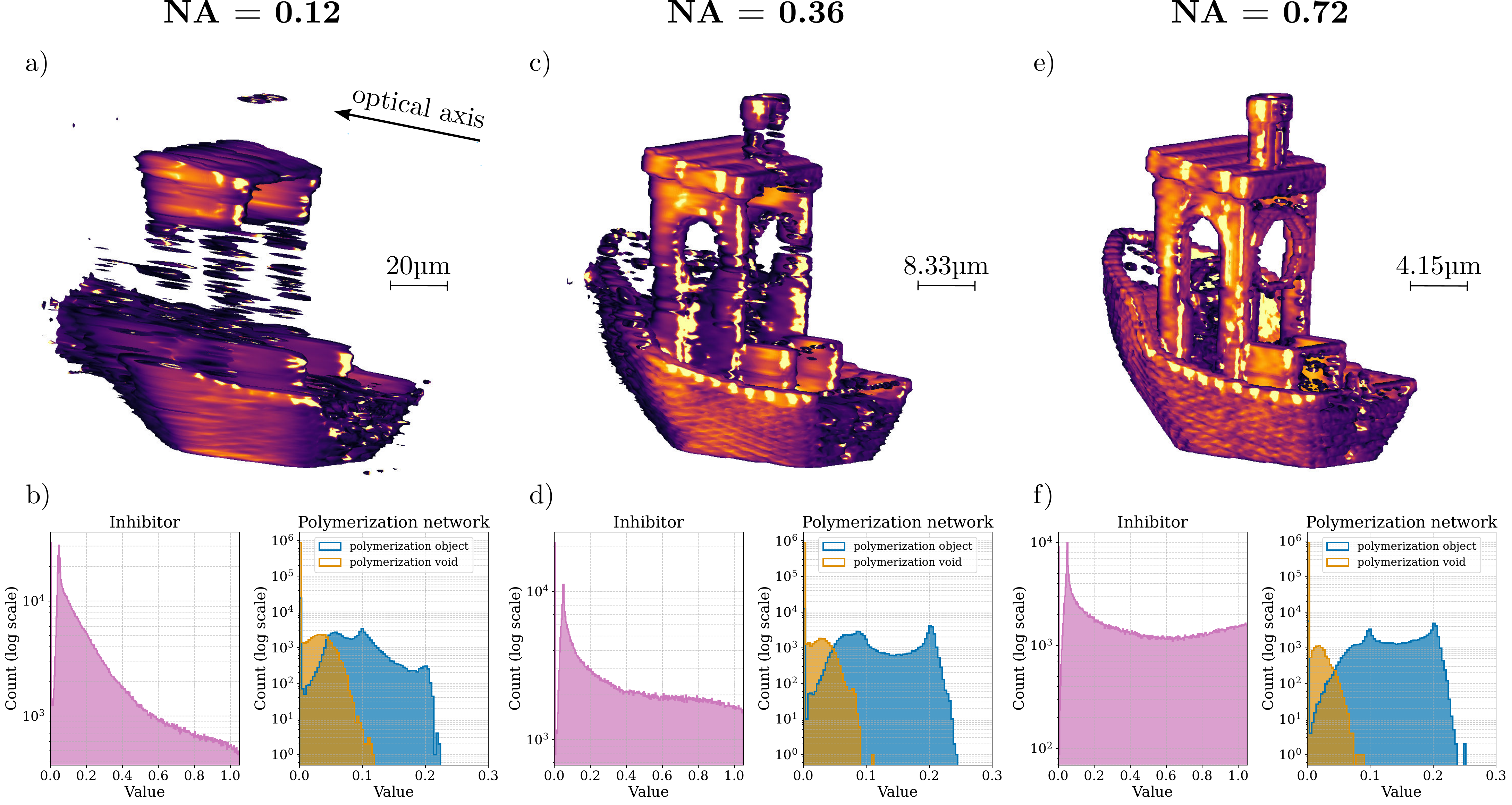}
    \caption{Simulated print outcomes for varying Numerical Apertures (NA). 
    \textbf{a, b)} Result and dose histogram for $\text{NA}=0.12$; the object is barely recognizable due to poor axial confinement. 
    \textbf{c, d)} Result and histogram for $\text{NA}=0.36$; volumetric features begin to resolve. 
    \textbf{e, f)} Result and histogram for $\text{NA}=0.72$; the geometry is reproduced with high fidelity. 
    Note that the physical field of view scales inversely with NA.}
    \label{fig:highna}
\end{figure*}

To ensure a fair comparison, we maintain a constant computational grid size of $100 \times 100 \times 100$ voxels across all simulations while scaling the physical field of view inversely with the NA. This reflects the practical trade-off in microscopy where higher resolution typically necessitates a smaller field of view. We evaluate three configurations: our experimental baseline of $\text{NA}=0.12$ (pixel size $\SI{1.69}{\micro\meter}$), a moderate $\text{NA}=0.36$ (pixel size $\SI{0.56}{\micro\meter}$), and a high $\text{NA}=0.72$ (pixel size $\SI{0.28}{\micro\meter}$). In all cases, we optimize a sequence of 30 time-multiplexed phase patterns.

The results are summarized in \autoref{fig:highna}. At $\text{NA}=0.12$ (\autoref{fig:highna}a), the reconstruction is severely degraded. The limited angular diversity results in a large missing cone in the frequency domain, causing elongation of features along the optical axis and preventing the successful resolution of the boat's hull and cabin.

Increasing the $\text{NA}$ to 0.36 (\autoref{fig:highna}c) significantly improves axial confinement, with the boat geometry becoming clearly recognizable. At $\text{NA}=0.72$ (\autoref{fig:highna}e), the system achieves true volumetric addressing; the complex overhangs and internal voids of the \#3DBenchy are almost fully reproduced. This improvement is quantitatively confirmed by the dose histograms (\autoref{fig:highna}b, d, f), which show a progressive sharpening of the contrast between object and void regions as NA increases.

These results indicate that SHVAM is capable of high-fidelity volumetric printing given sufficient optical access. While we simulated up to $\text{NA}=0.72$, immersion objectives could theoretically extend effective NAs to $\approx 1.4$, further minimizing the missing cone. We note that achieving this resolution experimentally would require accounting for chemical effects (as we presented in this work); as diffusion would blur these fine features, such high-NA implementations would likely necessitate diffusion-engineered resins or reduced print times.

%end appendix
}

\end{document}